\documentclass[twocolumn,showpacs,showkeys,preprintnumbers,amsmath,amssymb,prd,letterpaper,floatfix,superscriptaddress,10pt,aps,nofootinbib]{revtex4-1}
\usepackage[top=25truemm,bottom=25truemm,left=25truemm,right=25truemm]{geometry}
\usepackage{bbm,bm,graphicx,mathtools,color,slashed}
\usepackage{amssymb,amsmath}
\usepackage[export]{adjustbox}
\usepackage[colorlinks=true, linkcolor=blue, citecolor=blue, urlcolor=blue]{hyperref}
\usepackage[T1]{fontenc}
\usepackage{youngtab}
\usepackage{tabularx}
\usepackage{url}

\usepackage{tikz}
\usetikzlibrary{quantikz}
\usepackage{rotating}

\usepackage{mathtools}
\usepackage{cleveref}
\usepackage{placeins}
\usepackage{xfrac}
\usepackage{xcolor}

\allowdisplaybreaks

\newcommand{\tp}{\otimes}

\def\beq{\begin{equation}}
\def\eeq{\end{equation}}
\def\bea{\begin{eqnarray}}
\def\eea{\end{eqnarray}}
\def\barr{\begin{array}}
\def\earr{\end{array}}

\def\BNL{Department of Physics, Brookhaven National Laboratory, Upton, NY 11973, USA}
\def\RIKEN{RIKEN BNL Research Center, Brookhaven National Laboratory, Upton, NY 11973, USA}

\begin{document}

\title{Effects of Cosine Tapering Window on Quantum Phase Estimation}

\author{Gumaro Rendon}\email{jrendonsu@bnl.gov}
\affiliation{\BNL}

\author{Taku Izubuchi}
  \affiliation{\BNL}
  \affiliation{\RIKEN}

\author{Yuta Kikuchi}
  \affiliation{\BNL} 

\begin{abstract}
We provide a modification to the quantum phase estimation algorithm (QPEA)\cite{Abrams_QPE,Cleve_1998,nielsen2002quantum} inspired on classical windowing methods for spectral density estimation. From this modification we obtain an upper bound in the cost that implies a cubic improvement with respect to the algorithm's error rate. Numerical evaluation of the costs also demonstrates an improvement. Moreover, with similar techniques, we detail an iterative projective measurement method for ground state preparation that gives an exponential improvement over previous bounds using QPEA. Numerical tests that confirm the expected scaling behavior are also obtained. For these numerical tests we have used a Lattice Thirring model as testing ground.
Using well-known perturbation theory results, we also show how to more appropriately estimate the cost scaling with respect to state error instead of evolution operator error.
\end{abstract}

\maketitle

\section{Introduction}

In recent years, there has been increased interest in quantum computing techniques from the high-energy and nuclear physics community. Traditionally, physical systems that cannot be accessed through perturbative methods are studied through a combination of a Wick rotation and Monte Carlo methods~\cite{Gattringer:2010zz} to perform observables estimation in the Lagrangian formalism. This approach amounts to estimating the path integral in Euclidean space time. Working in Euclidean space time ensures the integrand, which is the Boltzmann weight used to sample Field configurations, is strictly positive and thus can be assigned to the probabilistic weight in Monte Carlo methods. However, this excludes systems that have topological terms~\cite{Izubuchi:2008mu}, non-zero chemical potentials~\cite{Aarts:2015tyj}, or limits the study of real-time dynamics in general~\cite{Alexandru:2016gsd,Takeda:2019idb}. These particular cases spoil the positiveness of the sampling weight and can no longer use the probabilistic interpretation.
Switching altogether to the Hamiltonian formalism avoids the requirement of probabilistic interpretation of Monte Carlo methods. The main drawback of working in the Hamiltonian formalism is the exponential growth of the Hilbert space dimension with respect to the system size. This renders most classical methods unfeasible even at relatively small system sizes. Quantum computing promises to bypass this problem~\cite{feynman1982simulating} in a general sense. Thus, it will be of importance to further develop quantum algorithms so we are ready on the arrival of proper quantum hardware.

One of the most relevant algorithms in the repertory of quantum computing is the quantum phase estimation algorithm (QPEA). An idea for quantum phase estimation algorithm first appeared in~\cite{kitaev1995quantum} as a method that infers the digit of a phase, one after the other, from least to most significant. Later came the version that relies on the quantum Fourier transform (QFT)~\cite{Abrams_QPE}, which was further detailed in~\cite{Cleve_1998}. 
QPEA has become an essential component of a lot of other algorithms, such as Shor's algorithm for the factorization of prime numbers~\cite{shor1994algorithms}, and quantum amplitude estimation algorithm~\cite{2000quant.ph..5055B}. Another possible use of the QPEA is in quantum state preparation~\cite{kitaev1995quantum}. In this context, the QPEA acts as a filter to
quantum states after measuring the ancillary register and obtaining the desired outcome. In that same vein, there is the Harrow-Hassidim-Lloyd (HHL) algorithm for solving the Quantum Linear System Problem~\cite{Harrow_2009}. This last algorithm was exponentially improved in its scaling with the error~\cite{Childs_2017} by replacing the QPEA part with the Linear-Combination-of-Unitaries (LCU) algorithm~\cite{LCU_2012} and Oblivious Amplitude Amplification (OAA)~\cite{Childs_2017}. This is just an example of the splurge of new developments in recent years using block-encoding and amplitude amplification that improve on previous algorithms, sometimes exponentially.

Some of these algorithms incur into large overheads and might be prohibitive at early stages of fault-tolerant quantum computing. For example, in Refs.~\cite{sprep_Cirac,Lin_2020} authors use block-encoding for state preparation. We detail here an alternative using Trotterization and QPEA, whose minimum number of ancillary qubits scale as $O(\log{1/\Delta})$, where $\Delta$ is the lower bound on the spectral gap of a given quantum system. 

Here, in \Cref{sec:imp_sprep}, we detail an iterative approach 
involving projective measurements
that exponentially improves the cost scaling with respect to the state error over previous estimates~\cite{sprep_Cirac}. Moreover, we study the effects of using the modifications proposed in \Cref{sec:imp_QPEA} for state preparation. This modification to the time window is also seen in Ref~\cite{cosine_2006}, and we present here a circuit to prepare such state at a cost scaling linearly with the number of ancillary qubits. Finally, in \Cref{sec:results_sprep}, we show some state preparation tests using a lattice implementation of the Thirring model~\cite{Ba_uls_2019}, and provide the tools to estimate excited state contamination.

\section{Spectral Density Estimation and the Effects of Windowing}

When calculating a spectral density numerically, we are bounded by the resources available, both classically or quantum mechanically. We cannot calculate the continuous-time Fourier transform, defined
\begin{align}
\mathcal{F}(f)(q)=\int^{\infty}_{-\infty} f(x) e^{-2\pi i x q} dq,
\end{align}
of a signal over an infinite time domain. Be it a sound recording, DC Voltage signal, or the Hamiltonian evolution of a quantum system, we must choose a finite sample rate and time domain.

Classically, one method to estimate the spectral density is to use instead the discrete Fourier transform (DFT)~\cite{Harris78onthe,blackman1958measurement}. It is defined as follows:
\begin{align}
  F_k &= \sum_{n=0}^{N-1} f_n \cdot e^{-\frac {i 2\pi}{N}kn},
\end{align}
where $\{f_n\}_n$ is a set of discrete samples of a continuous signal.

There are two effects to consider when estimating the spectral density from the discrete and finite samples of a signal. The two effects are aliasing, and spectral-leakage.

Aliasing effects can be understood as frequency-domain truncation effects (time discretization effects). When the frequency domain is not large enough to contain the whole spectrum, the spectrum outside the domain folds on top of the contained domain, producing the wrong spectral density. When the spectrum is discrete, one obtains Dirac deltas situated at the wrong energy values, thus the name aliasing effect. It is sufficient to extend the frequency domain (or, equivalently, increase the sampling rate) large enough to accommodate the whole spectrum ( by Cardinal theorem of interpolation~\cite{whittaker_e_t_1915_1428702,inbook,Shannon_sampling} ).

Spectral leakage, on the other hand, happens when we truncate the time domain. From all the possible frequencies in the continuum, only those that are exactly periodic on the interval will project to a single basis vector. This is the origin effect of spectral leakage~\cite{Harris78onthe}. It is the truncation of the time domain that causes the spectral leakage, not the finite sampling rate.

Typically, the Quantum Phase Estimation Algorithm (QPEA)~\cite{kitaev1995quantum,Abrams_QPE,Cleve_1998,nielsen2002quantum} consists of estimating the spectral density of the eigenstate $\ket{\psi_i}$ of $H$, i.e., $H\ket{\psi_i}=E_i\ket{\psi_i}$.
The frequency $E_i$ is what we attempt to estimate, that is, the corresponding eigenvalue to said eigenvector.

On the QPEA, as detailed in Refs.~\cite{Abrams_QPE,Cleve_1998,nielsen2002quantum}, we have two registers: one ancillary register and the target register. The target register stores the state $\ket{\psi_i}$ and the ancillary register will store the samples of the Hamiltonian evolution. The size of the ancillary register is $m$-qubit long and able to store a vector of size $2^m$. The size of the target register is $\log_{2}{\mathcal{H}}$, where $\mathcal{H}$ is the Hilbert space spanned by the eigenbasis $\{\ket{\psi_i}\}_i$.

The algorithm assumes we start with the state

\begin{align}
 \ket{0}_a \tp \vert \psi_i \rangle,
\end{align}
where $\ket{k}_{a}:=\ket{k_{m-1}}\otimes \ket{k_{m-2}}\otimes\cdots\otimes\ket{k_0}$ with $k_i\in\{0,1\}$ and $k=k_{m-1}2^{m-1} + k_{m-2}2^{m-2}+\dots+k_02^{0}$. We leave the target register unlabeled.

\begin{figure}[!htb]
\begin{quantikz}
\lstick[wires=5]{$\ket{0}^{\tp m}$}\slice{Spectral\\Filter} & \qw & \gate[wires=5,nwires=3]{QFT^{-1}}\slice[style={black}]{Time\\Window} &  \qw & \qw & \qw & \qw\ldots &\qw & \ctrl{5}\slice[style={blue}]{Samples $\times$ Window}  & \gate[wires=5,nwires=3]{QFT} & \meter{} & \rstick[wires=5]{$z$}
 \\
  & \qw &\qw &\qw& \qw&\qw & \qw\ldots & \ctrl{4}   & \qw & \qw & \meter{} &
 \\
  & \vdots & \vdots &\vdots& \vdots & \vdots & \vdots &\vdots&\vdots&\vdots&
 \\
  & \qw &\qw &\qw& \qw&\ctrl{2}  & \qw\ldots & \qw   & \qw & \qw & \meter{} &
 \\
  & \qw & \qw&\qw& \ctrl{1} & \qw  & \qw\ldots & \qw  & \qw  & \qw & \meter{} &
 \\
 \lstick{$\ket{\psi}$} &\qw \qw&\qw & \qw & \gate{U^{2^0}}& \gate{U^{2^1}} & \qw\ldots & \gate{U^{2^{m-2}}} & \gate{U^{2^{m-1}}} & \qw & \qw & 
\end{quantikz}
\caption{Circuit to implement an $m$-qubit phase estimation algorithm. $U$ is $e^{2 \pi i \lambda H}$. The $QFT^{-1}$ applied on $\ket{0}^{\tp m}$ can be simplified to $H^{\tp m}$ acting on $\ket{0}^{\tp m}$ as is typically shown. The red dashed line marks the samples of the filter function stored in the ancillary register and the black dashed line the samples of the corresponding window function.}\label{fig:mqubit_pea_textbook}
\end{figure}
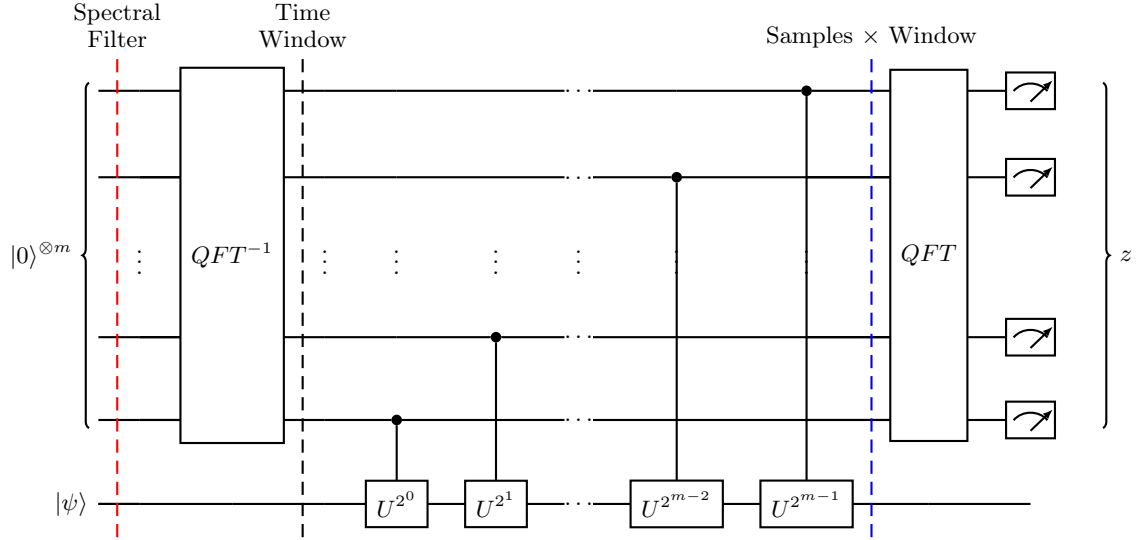

We can think of the controlled time evolution (the operations in between the black and blue dashed lines in \Cref{fig:mqubit_pea_textbook} ) in the QPEA as storing the finite (of size $2^m$) and discrete samples (taken at time intervals $2\pi\lambda$) of the time evolution in the ancillary register. However, the way the controlled evolution acts, we must take the ancillary register from $\ket{0}_a$ to a suitable window distribution first.

In the `textbook' version of the QPEA~\cite{Abrams_QPE,Cleve_1998,nielsen2002quantum} the window distribution used is the rectangular window, that is, a uniform superposition of all possible states. One way to achieve this is by applying the QFT$^{-1}$(IDFT, classically) to $\ket{0}_a$ (see \Cref{fig:mqubit_pea_textbook}, before the black dashed line) or, after a simplification of the circuit, an equivalent way to achieve this is applying Hadamard gates $H^{\tp m}$ on $\ket{0}_a$ (see \Cref{fig:simp_rectangular}).

Now, after having prepared the window distribution, and the controlled time evolution, what we have stored in the ancillary register are samples of the Hamiltonian evolution element-wise multiplied by the window amplitude (state at the blue dashed line in \Cref{fig:mqubit_pea_textbook}).

\begin{figure}[!htb]
\centering
\includegraphics[height=0.25\textwidth, valign=c]{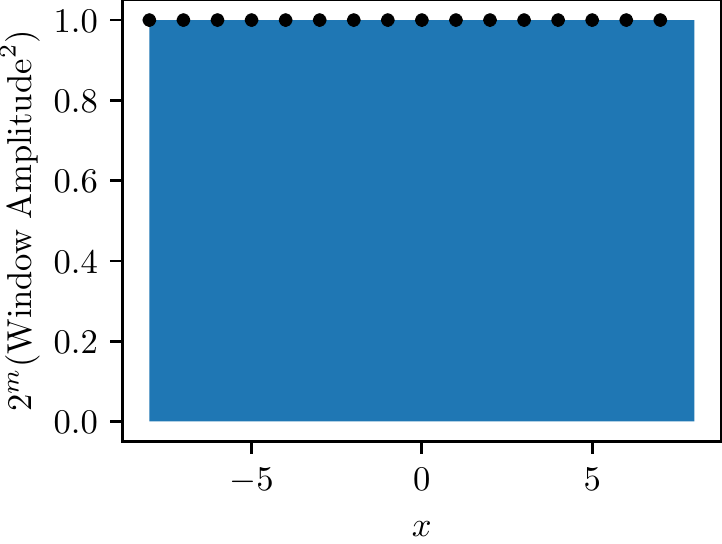} \( \quad \underset{QFT^{-1}}{\overset{QFT}{\rightleftarrows}} \quad \) \includegraphics[height=0.25\textwidth, valign=c]{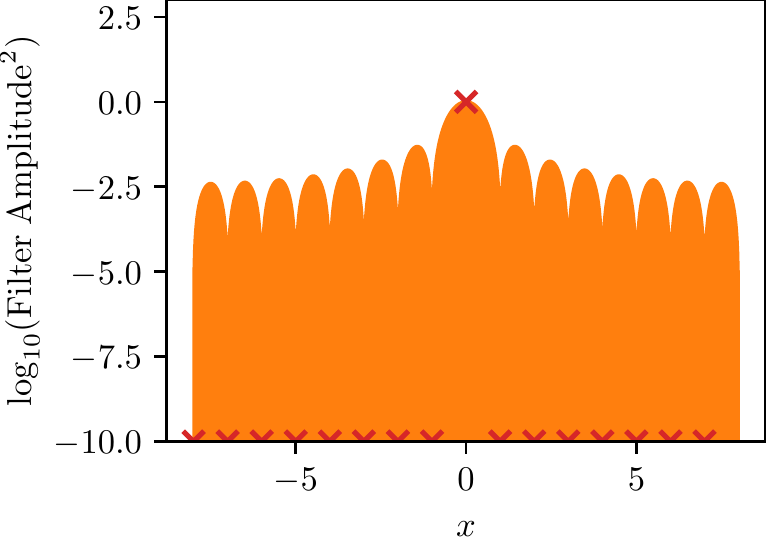} \\
\includegraphics[height=0.25\textwidth, valign=c]{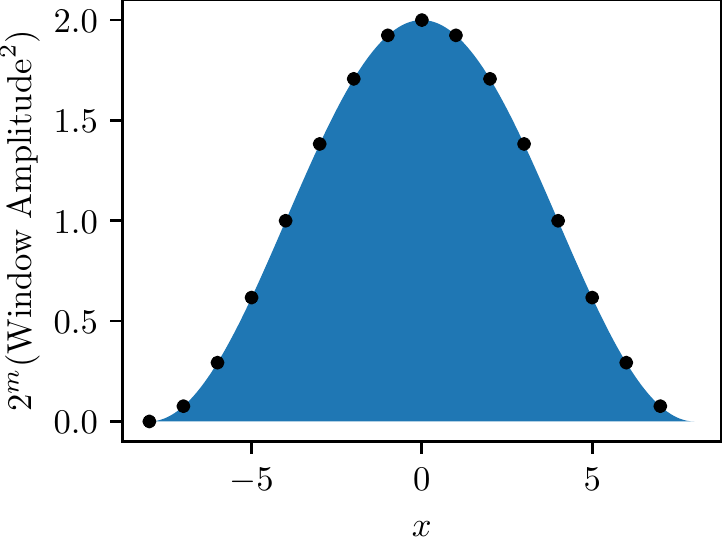} \( \quad \underset{QFT^{-1}}{\overset{QFT}{\rightleftarrows}} \quad \) \includegraphics[height=0.25\textwidth, valign=c]{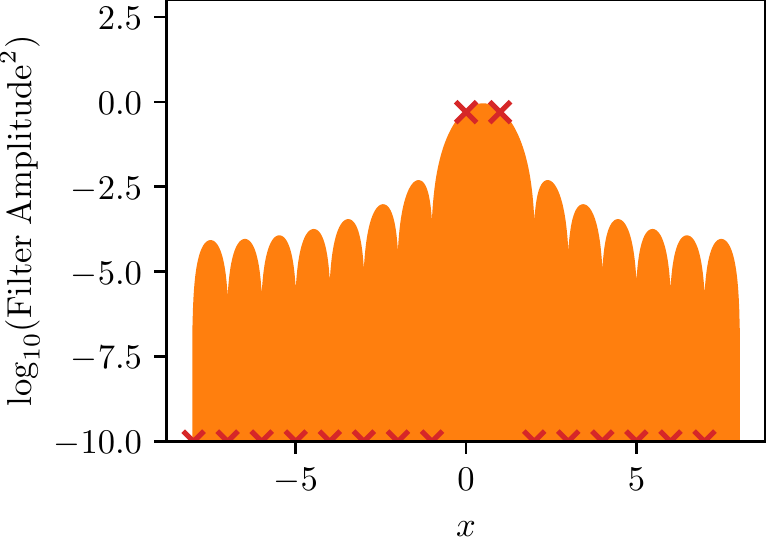} \\
\caption{Here we showcase the two different window functions and their corresponding filter functions, connected by DFT (QFT) and its inverse (QFT$^{-1}$). On the left, the black dots represent the discrete node points at which the window function is sampled. For example, these black dots represent the distribution of the ancillary register at the black dashed line on \Cref{fig:mqubit_pea_textbook} for the rectangular window and on \Cref{fig:mqubit_pea} for the cosine window. On the right, the red marks represent the distributions that we use to generate the windows through the inverse DFT (QFT) (these ancillary register distributions are marked with a red dashed line in \Cref{fig:mqubit_pea,fig:mqubit_pea_textbook} for the cosine and rectangular windows respectively).}\label{fig:new_vs_old_v2}
\end{figure}

Then, by applying the QFT operation on the ancillary register we are obtaining an approximate spectral density function of the state $\ket{\psi_i}$. The exact spectral density consists of a Dirac delta situated at the eigenvalue of $\ket{\psi_i}$. However, due to the finite time-domain effects (spectral leakage), what we get is a smeared-out version (See first row, second column of \Cref{fig:new_vs_old_v2}).

It is worth noting that, regardless of sign-in-the-exponential conventions, we have changed the last QFT$^{-1}$~\cite{Abrams_QPE,nielsen2002quantum,Cleve_1998} to a QFT because in  this context the purpose of the algorithm is to estimate the spectrum. Thus, we have switched the spaces and transforms from Refs.~\cite{Abrams_QPE,nielsen2002quantum,Cleve_1998}. Moreover, because of the sign-in-the-exponential conventions chosen here, the ``time-evolution'' operator, $U=e^{i2\pi\lambda H}$, has a positive sign in contrast to Schr\"odinger's equation.

One can also think of spectral leakage as coming from the discontinuities that arise from the periodization of a signal. If the signal contains frequencies that are not exactly periodic on the interval, discontinuities are introduced. Therefore, one could ameliorate this effect by multiplying the signal by a window that tapers off at the boundaries of the time-domain~\cite{Harris78onthe}.

One window that we can try is the cosine tapering window, which requires only a small modification to the original quantum phase estimation algorithm. We can compare both the rectangular window and the cosine window in \Cref{fig:new_vs_old_v2}. On the right column, we can see the corresponding DFT's (up to phasing operators $\sum_x e^{-\frac{2\pi i q_0 x}{2^m}}$ \ket{x}\bra{x} ). The black dots on the window and red marks on the filter functions denote the node positions and the evaluation of the functions at them. This gives a clue on how to generate the cosine tapering window.\\

\onecolumngrid

\twocolumngrid

\onecolumngrid

\begin{figure}[!htb]
\centering
\includegraphics[width=0.7\textwidth]{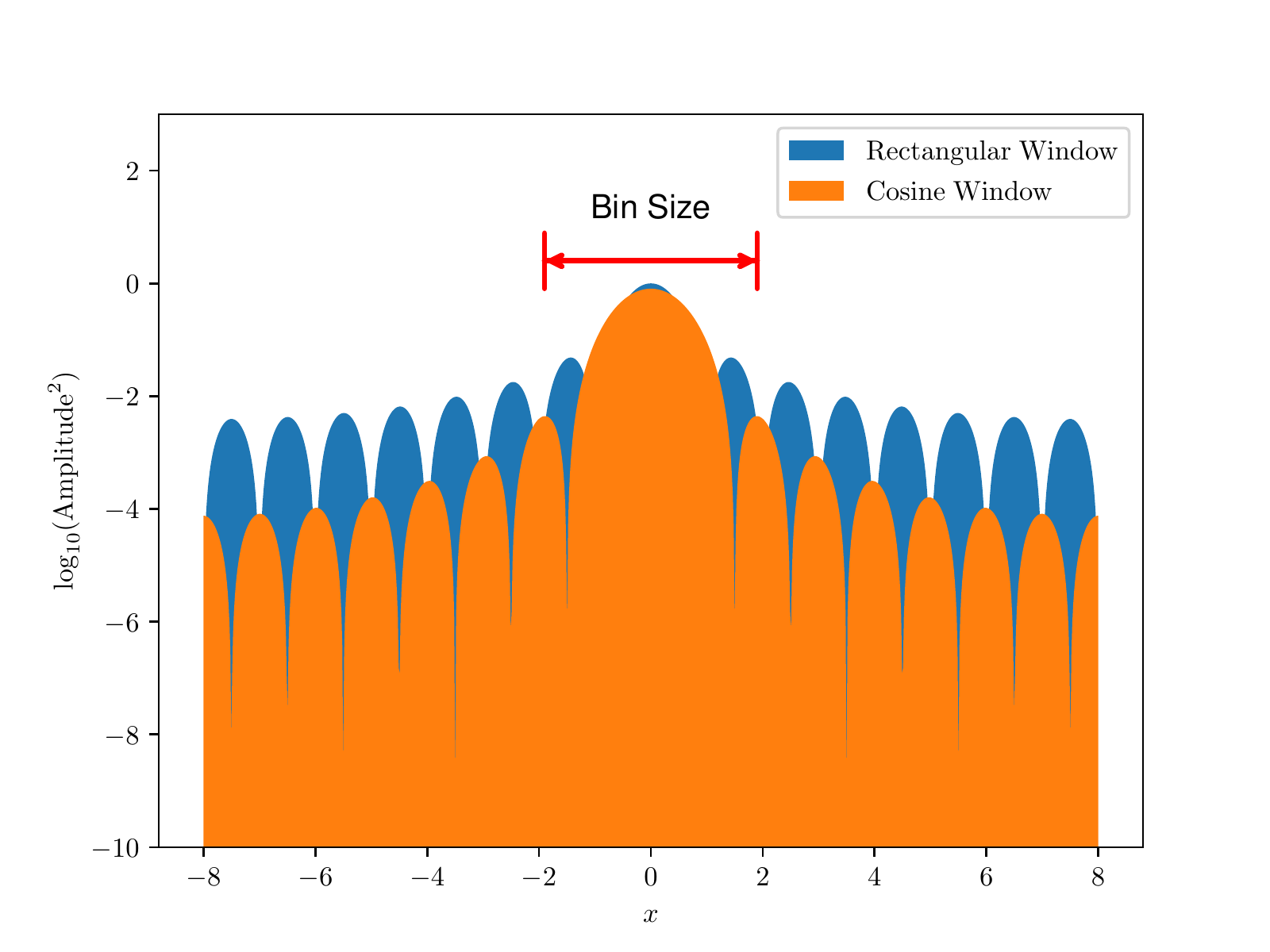}
\caption{Comparison of the filter functions in \Cref{fig:new_vs_old_v2}. We have shifted the cosine filter function in order for it to be centered at $q=0$. The filter corresponding to the cosine window has a wider main lobe, but it has a faster decaying behavior outside of that. As a consequence, when binning the probabilities, we get a higher peak for the cosine filter.}\label{fig:new_vs_old}
\end{figure}

\twocolumngrid

\onecolumngrid

\twocolumngrid


\section{Cosine Tapering Window in Quantum Phase Estimation\label{sec:imp_QPEA}}

\onecolumngrid

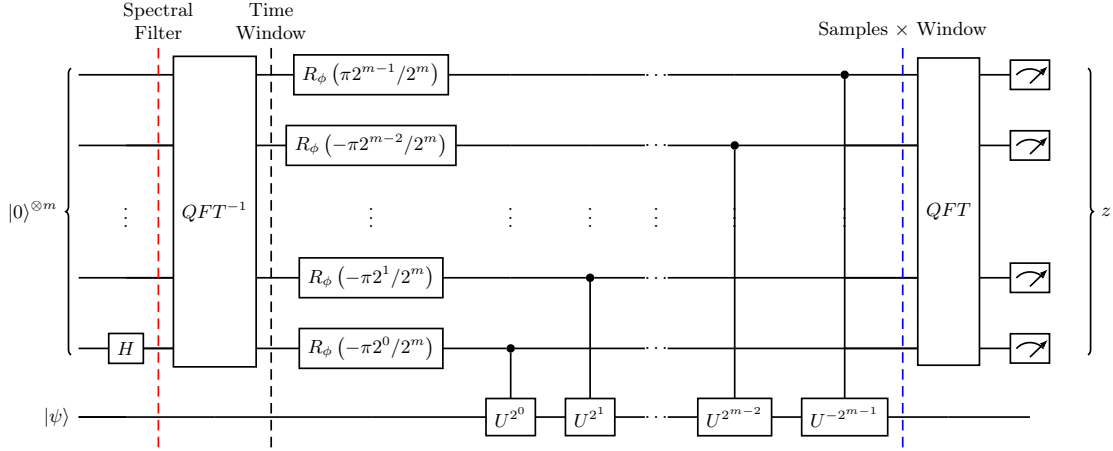
\begin{figure}[!htb]
\scalebox{0.8}{
\begin{quantikz}
\lstick[wires=5]{$\ket{0}^{\tp m}$} & \qw & \gate[wires=5,nwires=3]{QFT^{-1}}\slice[style={black}]{Time\\Window} &  \gate{R_\phi \left( \pi2^{m-1}/2^m \right)} & \qw & \qw & \qw\ldots &\qw & \ctrl{5}\slice[style={blue}]{Samples $\times$ Window}  & \gate[wires=5,nwires=3]{QFT} & \meter{} & \rstick[wires=5]{$z$}
 \\
  & \qw &\qw &\gate{R_\phi \left(-\pi2^{m-2}/2^m \right)}& \qw&\qw & \qw\ldots & \ctrl{4}   & \qw & \qw & \meter{} &
 \\
  & \vdots & \vdots &\vdots& \vdots & \vdots & \vdots &\vdots&\vdots&\vdots&
 \\
  & \qw &\qw &\gate{R_\phi \left( -\pi2^{1}/2^m \right)}& \qw&\ctrl{2}  & \qw\ldots & \qw   & \qw & \qw & \meter{} &
 \\
  & \gate{H}\slice{Spectral\\Filter}& \qw&\gate{R_\phi \left(-\pi2^{0}/2^m \right)}& \ctrl{1} & \qw  & \qw\ldots & \qw  & \qw  & \qw & \meter{} &
 \\
 \lstick{$\ket{\psi}$} &\qw \qw&\qw & \qw & \gate{U^{2^0}}& \gate{U^{2^1}} & \qw\ldots & \gate{U^{2^{m-2}}} & \gate{U^{-2^{m-1}}} & \qw & \qw & 
\end{quantikz}
}
\caption{Circuit to implement improved $m$-qubit phase estimation algorithm. $U$ is $e^{2 \pi i \lambda H}$. The red dashed line marks the samples of the filter function stored in the ancillary register and the black dashed line the samples of the corresponding window function.}\label{fig:mqubit_pea}
\end{figure}

\twocolumngrid

For the modified phase estimation with cosine tapering window, we again assume that we start with the state
\begin{align}
 \ket{0}_a \tp \vert \psi_i \rangle,
\end{align}
where the subscript $a$ indicates that the first register is the ancillary register. Since the subsequent quantum operations does not alter the state on the target register, we suppress $\ket{\psi_i}$ in what follows.
We first start by creating a super position of $\ket{0}_{a}$ and $\ket{1}_{a}$ on the ancillary register by applying a Hadamard gate on the least significant qubit
\begin{align}
 \frac{\ket{0}_a+\ket{1}_a}{\sqrt{2}}.
\end{align}
The inverse Quantum Fourier Transform (QFT$^{-1}$), in the positive-integers bases, is realized by the following operator:
\begin{align}
\frac{1}{\sqrt{2^m}}  \sum^{2^{m}-1}_{y=0} \sum^{2^{m}-1}_{k=0} e^{\frac{2 \pi i y k}{2^m}} \ket{y}\bra{k}.
\end{align}
For reasons that will become apparent in the next section, we work in the almost-centered bases. That is, we relabel the states
\begin{align}
\ket{y}\to 
\left\{
\begin{array}{ll}
\ket{x=y} & (0\leq y \leq2^m/2-1), 
\\
\ket{x=y-2^m} & (2^m/2\leq y \leq 2^m-1).
\end{array}
\right.
\end{align}
We have the same map from $k$ to $q$. The circuit for QFT$^{-1}$ need not change, as we just do a relabeling of states (see \Cref{appendix:change_of_bases}). In these new bases the operator becomes
\begin{align}
\frac{1}{\sqrt{2^m}} \sum^{2^{m-1}-1}_{x=-2^{m-1}} \sum^{2^{m-1}-1}_{q=-2^{m-1}}  e^{\frac{2 \pi i x q}{2^m}} \ket{x}\bra{q}
=: QFT^{-1}.
\end{align}
Now, we perform this QFT$^{-1}$ operation which leaves us with
\begin{align}
\begin{split}
 &QFT^{-1}\left(\frac{\ket{0}_a+\ket{1}_a}{\sqrt{2}}\right)
 =\sum^{2^{m-1}-1}_{x=-2^{m-1}} f(x) \vert x \rangle_a,
\end{split}
\end{align}
where
\begin{align}
f(x) = \frac{1+e^{\frac{2 i \pi  x}{2^m}}}{\sqrt{2^{m+1}}}.
\end{align}
That is not the cosine window just yet. In order to obtain the cosine window, we must apply the phase $e^{-\frac{i \pi  x}{2^m}}$. This is equivalent to centering the corresponding spectral filter seen on the lower right of \Cref{fig:new_vs_old_v2}. This can be done through applying
\begin{align}
    R_{\phi,m-1}\left(\frac{\pi 2^{m-1}}{2^m} \right)\otimes \bigotimes^{m-2}_{l=0} R_{\phi,l}\left(-\frac{\pi 2^{l}}{2^m} \right).
\end{align}
We are left with
\begin{align}\label{eq:ftilde_psi}
 \sum^{2^{m-1}-1}_{x=-2^{m-1}} \tilde{f}(x) \vert x \rangle_a,
\end{align}
where
\begin{align}\label{eq:cosine_window}
\tilde{f}(x) = f(x) \exp \left(-\frac{ \pi  i x}{2^{m}}\right) = \frac{ \sqrt{2} \cos \left(\frac{\pi  x}{2^m}\right)}{\sqrt{2^m}}.
\end{align}
Thus, we have obtained the desired time-domain window.

Performing the controlled evolution operations as displayed in \Cref{fig:mqubit_pea} we obtain
\begin{align}
 &\sum^{2^{m-1}-1}_{q=-2^{m-1}}  \tilde{f}(x) e^{\frac{i 2 \pi x \lambda E_i}{2^m}} \vert x \rangle_a.
\end{align}
Note that the input state $\ket{\psi}$, which is suppressed here, is the target register of these controlled operations.
Finally, we apply the almost-centered QFT,
\begin{align}
    QFT
    :=\frac{1}{\sqrt{2^m}}
    \sum^{2^{m-1}-1}_{q=-2^{m-1}}
    \sum^{2^{m-1}-1}_{x=-2^{m-1}} e^{\frac{-2 \pi i x q}{2^m}} \ket{q}\bra{x}.
\end{align}
to find,
\begin{align}
\begin{split}
 &QFT\sum^{2^{m-1}-1}_{x=-2^{m-1}} \tilde{f}(x)e^{\frac{i 2 \pi x \lambda E_i}{2^m}} \vert x \rangle_a
 \\
 &=\sum^{2^{m-1}-1}_{q=-2^{m-1}} F(q-2^m\lambda E_i) \vert q \rangle_a,
\end{split}
\end{align}
where the coefficient $F(q)$ takes the form,
\begin{align}
\label{Fq}
F(q)&=\frac{1}{\sqrt{2^m}}\sum^{2^{m-1}-1}_{x=-2^{m-1}}\tilde{f}[x] e^{\frac{-2 \pi i x q}{2^m}},
\end{align}
which is the almost-centered DFT of $f(x)$.
Thus, the probability of measuring each value of $q$ on the ancillary register is
\begin{align}
P(q)=\lvert F\left(q-2^m \theta_i\right)\rvert^2,
\end{align}
with $\theta_i:=\lambda E_i$.
We now approximate $2^m \theta_i$ to the nearest integer. That is, $2^m \theta_i = z + 2^m \delta$, where $z$ is the nearest integer to $2^m \theta_i$ and $\vert 2^m \delta  \vert \leq \frac{1}{2} $.

We find that the lowest probability of measuring $z$ on the ancillary register is when $\vert 2^m \delta  \vert = \frac{1}{2}$ and that corresponds to $\min_\delta\text{Pr}(z) = \min_\delta\lvert F\left(-2^m \delta\right) \rvert^2 = \frac{1}{2}$. Therefore,
\begin{align}
\text{Pr}(z) \geq \frac{1}{2}.
\end{align}
That is an improvement of worst-case probability from $\text{Pr}(z)\geq \frac{4}{\pi^2}$ from the phase estimation in \cite{Abrams_QPE,Cleve_1998,nielsen2002quantum}.

In order to amplify the probability of success of obtaining an estimate of $\theta_i$, we must sacrifice in precision or cost~\cite{Cleve_1998}.
This can be understood as coarsening the data by summing the probabilities of $k=2^{p-1}$ results to the left and to the right (including $z$) of the nearest integer, $z$. That is,
\begin{align}
P(-k\leq l<k) = \sum_{z-k\leq l<z+k} \lvert \alpha_{lz} \rvert^2,
\end{align}
where
\begin{align}\label{eq:alpha_lz}
\alpha_{lz}=F(l-\delta 2^m).
\end{align}
Equivalently, we can define the%
complementary 
error rate $e=1-P(-k\leq l<k)$, which is 
\begin{align}\label{eq:error_rate}
e=\sum_{k\leq l < 2^{m-1} } \lvert \alpha_{lz} \rvert^2 + \sum_{-2^{m-1}\leq l < -k } \lvert \alpha_{lz} \rvert^2.
\end{align}
Summing up the probabilities (coarsening the data) has the effect of raising the tolerance in precision from $1/2^{m+1}$ to $k/2^{m}=1/2^{m-p+1}$. Therefore, it is convenient to define a new variable $t$ through the following
\begin{align}\label{eq:mtp}
m=t+p,
\end{align}
such that the target precision to which we estimate the phase $\theta_i$ is $1/2^{t+1}$. This way the number of qubits $t$ determines the target precision and the number of extra qubits $p$ determines the probability of failure of the method, or error rate, defined by $e$.

We can obtain an upper bound on $e$ as done in Refs.~\cite{Cleve_1998,nielsen2002quantum}. In these references, authors obtain this upper bound on the error rate, but for the circuit in \Cref{fig:mqubit_pea_textbook} ( with $\alpha_{lz}$ corresponding to rectangular window). With this, they then solve for the number of summed-up qubits $p_{\text{rect}}$, and obtain that the minimum requirement for a certain target rate $e$ is
\begin{align}\label{eq:p_rect}
p_{\text{rect}}=\left\lceil \log_2{\frac{1}{2e}+\frac12} \right\rceil
\end{align}

In \Cref{appendix:qubit_bound}, we derive an analogous upper bound on $e$ for the cosine-window filter from which we obtain the minimum number of extra qubits
\begin{align}
p=\left\lceil\log_2\left(\frac{\pi^{2/3}}{48^{1/3} e^{1/3}}+2\right)\right\rceil
\end{align}
for a target error rate $e$.

The cost in the gate complexity increases multiplicatively by $2^p$. 
This means a constant improvement in the minimum bound for $p$, but a cubic improvement in gate complexity compared to Refs.~\cite{Abrams_QPE,Cleve_1998,nielsen2002quantum} with respect to $e$. This can be also seen from the shape of both filters superposed in \Cref{fig:new_vs_old}, the amplitude appears to be more concentrated for the cosine-window filter than for the rectangular-window one.

\begin{figure}[!htb]
\centering
\includegraphics[width=0.7\textwidth]{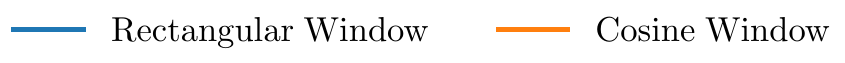}\\
\includegraphics[width=0.45\textwidth]{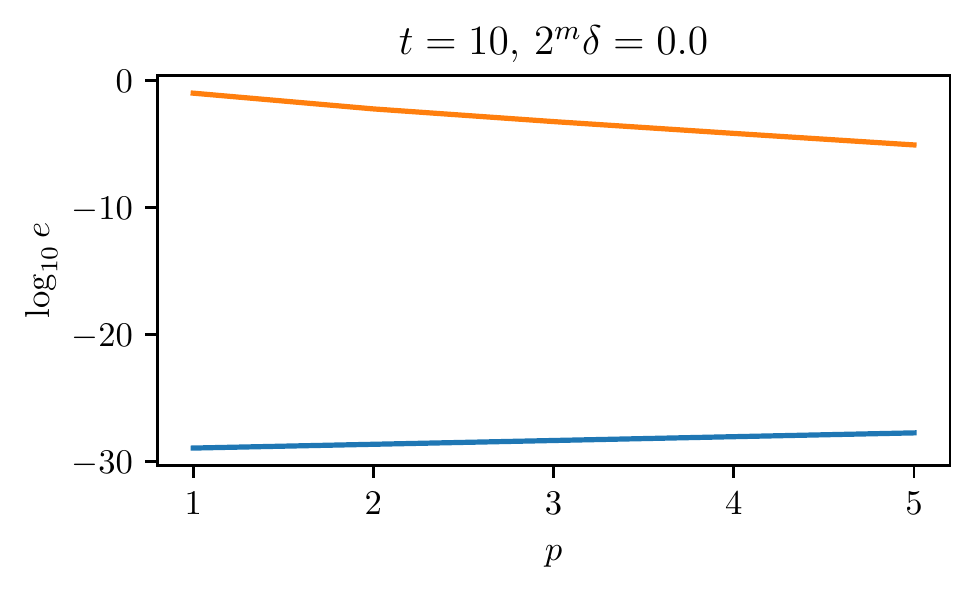} \includegraphics[width=0.45\textwidth]{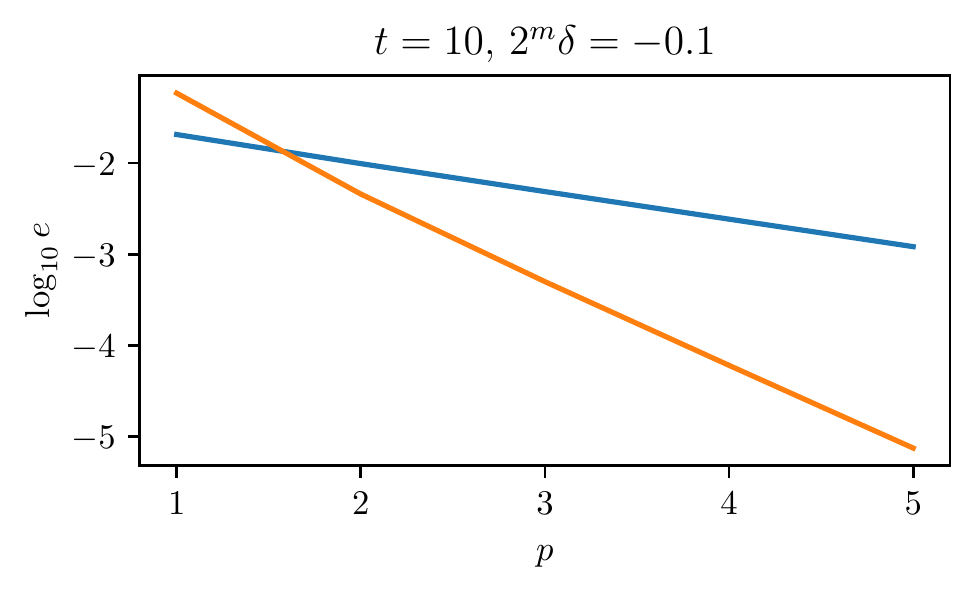} \\
\includegraphics[width=0.45\textwidth]{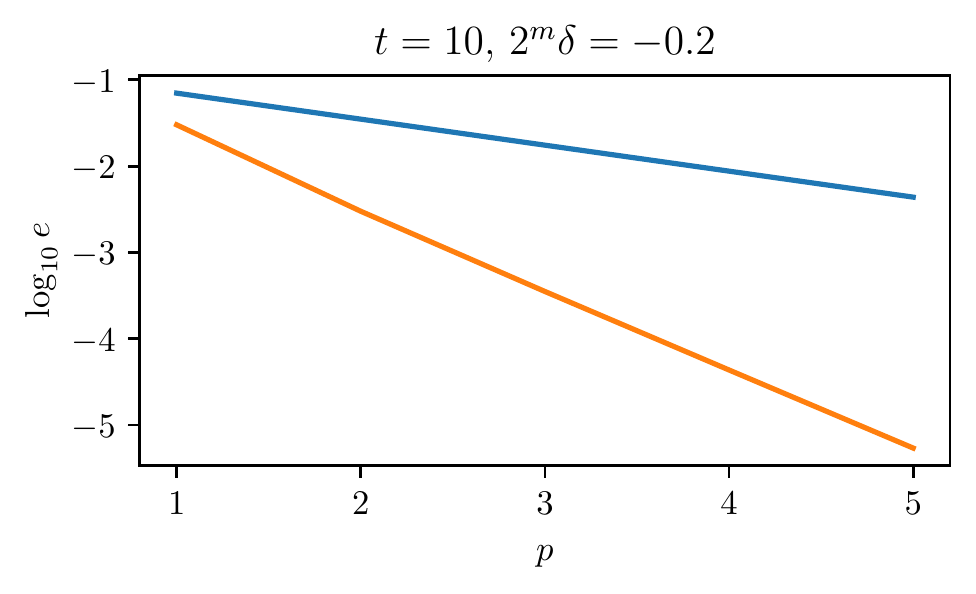} \includegraphics[width=0.45\textwidth]{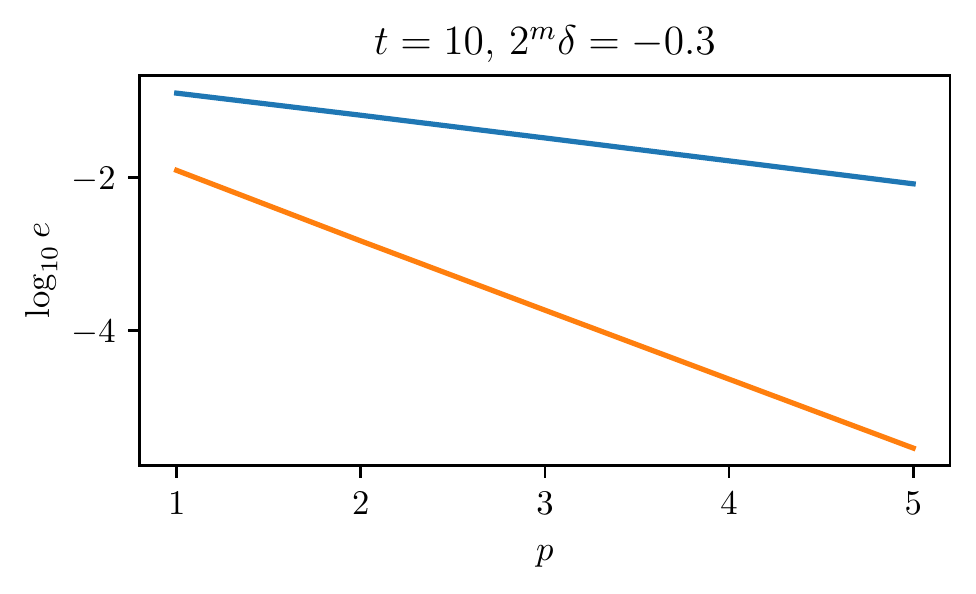} \\
\includegraphics[width=0.45\textwidth]{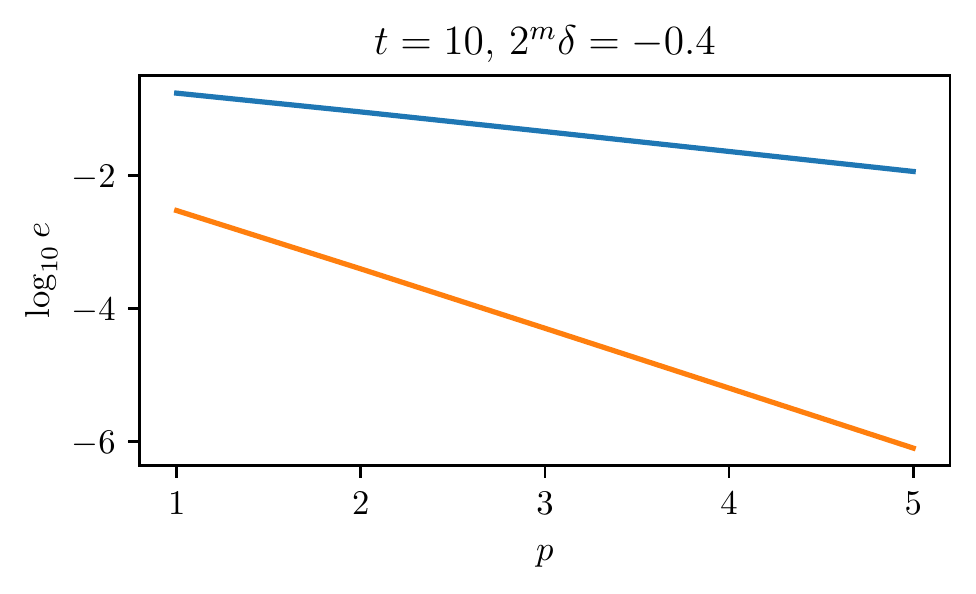} \includegraphics[width=0.45\textwidth]{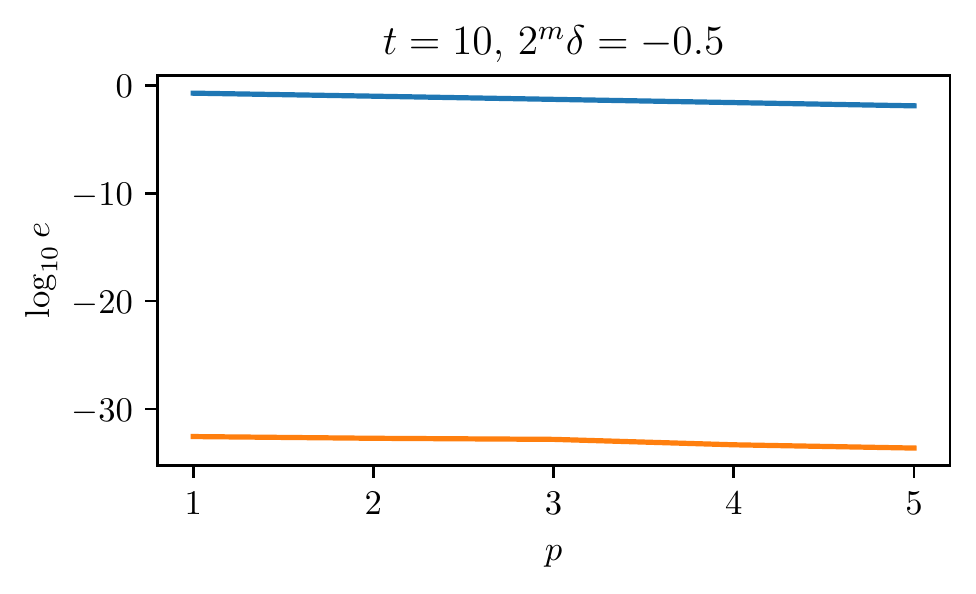}
\caption{Here we plot the error rate $e$ defined in \Cref{eq:error_rate} against the extra number of qubits $p$ in \Cref{eq:mtp}. We have chosen $t=10$ and $2^m\delta\in\{0.0,-0.1,-0.2,-0.3,-0.4,-0.5\}$. It is clear that, except for the exceptional cases $2^m\delta=0$, the filter function coming from the cosine window outperforms the one from the rectangular window as we increase $p$. \label{fig:e_vs_p}}
\end{figure}

Since we have only obtained a bound, we also perform numerical tests and plot $e$ against $p$ for $m=10$ on \Cref{fig:e_vs_p} for different values of $2^m\delta$. The cases $2^m\delta=0$ and $-0.5$ are exceptional because we expect the error rate to be zero for the rectangular-window filter for $2^m\delta=0$ for any value of $p$, and zero for the cosine-window filter for $\lvert 2^m\delta\rvert=0.5$ and for $p\geq1$. In the corresponding plots in \Cref{fig:e_vs_p} we see that those cases expected to be zero are of the order $\sim 10^{-30}$ due to numerical precision. We can see that the cosine-window filter outperforms the rectangular-window filter as we increase $p$, except for the particular case $2^m\delta=0$.

We also note that this particular window was also obtained previously~\cite{cosine_2006} as the choice that optimizes the alternative variance definition for discrete variable distributions:
\begin{align}
\bar{C}=\frac{1}{2\pi}\sum_z \int^{2\pi}_{\theta=0} d\theta \Pr(z|\theta) \sin^2\left(\frac{\theta - \frac{2\pi z}{2^m}}{2}\right).
\end{align}
In this section, we have put this same window to the test under another metric: algorithmic error rate for phase estimation. In the following section, we will estimate the circuit depth for state preparation using the rectangular-window filter as well as the cosine-window filter in order to compare.\\

\clearpage

\onecolumngrid

\twocolumngrid

\section{Iterative State Preparation and the Effects of Cosine Tapering Window\label{sec:imp_sprep}}

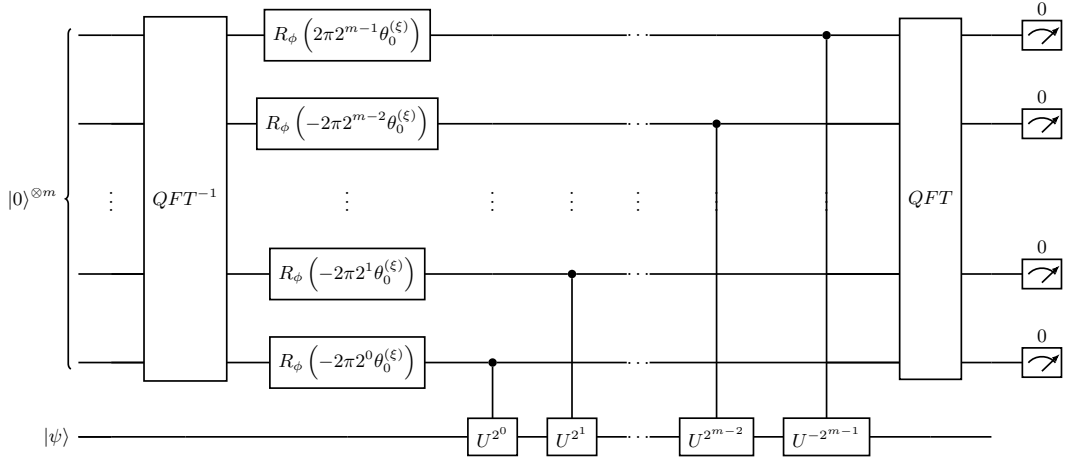
\begin{figure}[!htb]
\centering
\scalebox{0.8}{
\begin{quantikz}
\lstick[wires=5]{$\ket{0}^{\tp m}$} & \qw & \gate[wires=5,nwires=3]{QFT^{-1}} &  \gate{R_\phi \left(2\pi2^{m-1}\theta^{(\xi)}_{0} \right)} & \qw & \qw & \qw\ldots &\qw & \ctrl{5}  & \gate[wires=5,nwires=3]{QFT} & \qw & \meter{0}
 \\
  & \qw &\qw &\gate{R_\phi \left(-2\pi2^{m-2}\theta^{(\xi)}_{0} \right)}& \qw&\qw & \qw\ldots & \ctrl{4}   & \qw & \qw & \qw & \meter{0}
 \\
  & \vdots & \vdots &\vdots& \vdots & \vdots & \vdots &\vdots&\vdots&\vdots&
 \\
  & \qw &\qw &\gate{R_\phi \left(-2\pi2^{1}\theta^{(\xi)}_{0} \right)}& \qw&\ctrl{2}  & \qw\ldots & \qw   & \qw & \qw & \qw & \meter{0}
 \\
  & \qw& \qw&\gate{R_\phi \left(-2\pi2^{0}\theta^{(\xi)}_{0} \right)}& \ctrl{1} & \qw  & \qw\ldots & \qw  & \qw  & \qw & \qw & \meter{0}
 \\
 \lstick{$\ket{\psi}$} &\qw \qw&\qw & \qw & \gate{U^{2^0}}& \gate{U^{2^1}} & \qw\ldots & \gate{U^{2^{m-2}}} & \gate{U^{-2^{m-1}}} & \qw & \qw & 
\end{quantikz}
}
\caption{Circuit to implement projection with peak centered at the estimate of the ground state energy. The resulting filter is the DFT of a rectangular window or a constant signal.}\label{fig:iteration_rectangular}
\end{figure}

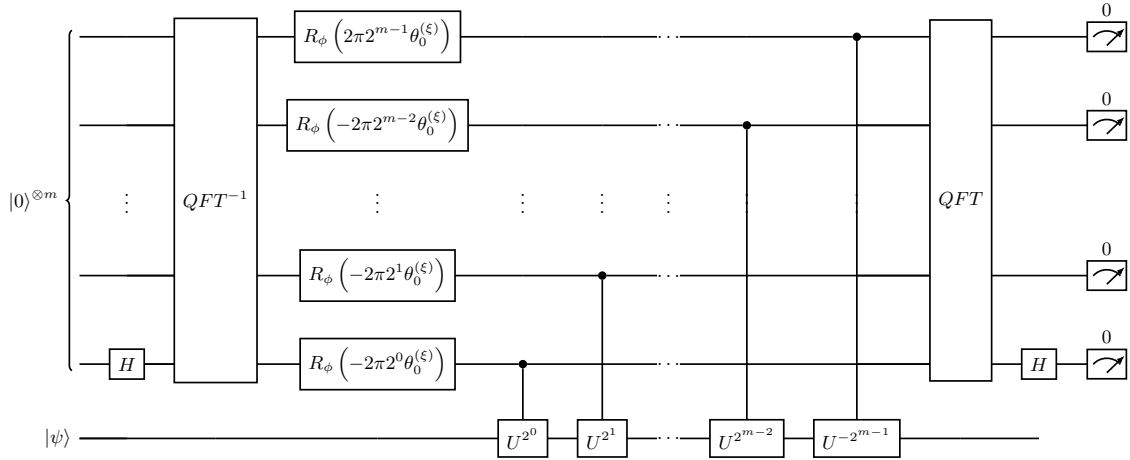
\begin{figure}[!htb]
\centering
\scalebox{0.8}{
\begin{quantikz}
\lstick[wires=5]{$\ket{0}^{\tp m}$} & \qw & \gate[wires=5,nwires=3]{QFT^{-1}} &  \gate{R_\phi \left(2\pi2^{m-1}\theta^{(\xi)}_{0} \right)} & \qw & \qw & \qw\ldots &\qw & \ctrl{5}  & \gate[wires=5,nwires=3]{QFT} & \qw & \meter{0}
 \\
  & \qw &\qw &\gate{R_\phi \left(-2\pi2^{m-2}\theta^{(\xi)}_{0} \right)}& \qw&\qw & \qw\ldots & \ctrl{4}   & \qw & \qw & \qw & \meter{0}
 \\
  & \vdots & \vdots &\vdots& \vdots & \vdots & \vdots &\vdots&\vdots&\vdots&
 \\
  & \qw &\qw &\gate{R_\phi \left(-2\pi2^{1}\theta^{(\xi)}_{0} \right)}& \qw&\ctrl{2}  & \qw\ldots & \qw   & \qw & \qw & \qw & \meter{0}
 \\
  & \gate{H}& \qw&\gate{R_\phi \left(-2\pi2^{0}\theta^{(\xi)}_{0} \right)}& \ctrl{1} & \qw  & \qw\ldots & \qw  & \qw  & \qw & \gate{H} & \meter{0}
 \\
 \lstick{$\ket{\psi}$} &\qw \qw&\qw & \qw & \gate{U^{2^0}}& \gate{U^{2^1}} & \qw\ldots & \gate{U^{2^{m-2}}} & \gate{U^{-2^{m-1}}} & \qw & \qw & 
\end{quantikz}
}
\caption{Circuit to implement projection with peak centered at the estimate of the ground state energy. The resulting filter is the DFT of a cosine tapering window. The last Hadamard gate is there to create a coherent binning of odd and even $x$ states.}\label{fig:iteration_cosine}
\end{figure}

The phase estimation algorithm can also be used for state preparation. Here, we will detail how that is accomplished. First, using the rectangular-window variation (\Cref{fig:iteration_rectangular}), and finally comparing with our cosine-window version (\Cref{fig:iteration_cosine}). First, we assume the target register starts with the state
\begin{align}\label{eq:init_state}
\ket{\phi} = \sum_i \phi_i \ket{\psi_i},
\end{align}
where $\ket{\psi_i}$ are the eigenstates of the Hamiltonian in question and $\phi_i$ are the overlap factors of the initial guess with those eigenstates.

In the broad sense, the state preparation method detailed here consists of applying the phase estimation circuit in \Cref{fig:iteration_rectangular} on the state $\ket{0}_a\ket{\phi}$. This is done multiple times until the resulting state, $\ket{\psi}$, is $\epsilon$-close to the ground state $\ket{\psi_0}$,
\begin{align}\label{eq:epsilon1}
\lVert \ket{\psi} - \ket{\psi_0} \rVert \le \epsilon.
\end{align}
It is only through using this iterative approach that we achieve a logarithmic cost scaling with respect to $1/\epsilon$ as explained below.

The phase estimation algorithm in \Cref{fig:iteration_rectangular} is an approximate implementation of the projection operator on the ground state. We assume here that we can simulate the Hamiltonian evolution exactly.
Therefore, applying the phase estimation circuit \Cref{fig:iteration_rectangular}, just before the measurement on $\ket{0}_a \ket{\phi}$, leaves us with the state 
\begin{align}\label{eq:gx_before}
\sum_i\sum^{2^{m-1}-1}_{q=-2^{m-1}} G\left(q-(2^m \theta_i - 2^m \theta^{(\xi)}_0)\right) \phi_i \vert q \rangle_a \tp \vert \psi_i \rangle,
\end{align}
which up to this point is a unitary operation. In the last expression, $\theta_i$ represents $\lambda E_i$ and $\theta^{(\xi)}_0$ is our best estimate of $\theta_0=\lambda E_0$, where $\xi$ is the precision to which we know $\theta_0$. The norm is preserved by Parseval's theorem for the DFT, that is, $\sum_q \lvert G\left(q-x\right) \rvert^2=1$ for any $x$. Here, we have used $G(q)$ corresponding to the rectangular window instead of $F(q)$ for the cosine window. We can check that the rectangular-window filter takes the form,
\begin{align}\label{eq:gx}
G(q)=\frac{e^{\frac{i \pi  q}{2^m}} \sin (\pi  q)}{2^m  \sin \left(\frac{\pi  q}{2^m}\right)}=e^{\frac{i \pi  q}{2^m}} D_{2^m}\left(\frac{q}{2^m}\right).
\end{align}
The $R_\phi$ gates in \Cref{fig:iteration_rectangular} provide the shift $\theta^{(\xi)}_0$ to the filter function in \Cref{eq:gx_before}.
\onecolumngrid

\twocolumngrid
Measuring the ancillary register and post-selecting $\ket{0}_a$ results in the state,
\begin{align}
\sum_i G\left( 2^m \theta^{(\xi)}_0-2^m \theta_i\right) \phi_i \vert 0 \rangle_a \tp \vert \psi_i \rangle.
\end{align}
The whole circuit (with measurement included) is equivalent to applying the following filter or approximate projector on the target register
\begin{align}\label{eq:psi0_filter}
\tilde{P}_{\psi_0} = \sum_i \gamma_i \ket{\psi_i}\bra{\psi_i},
\end{align}
where $\gamma_i=G\left(2^m \theta^{(\xi)}_0-2^m \theta_i \right)$. We can also recast the approximate projector to the form
\begin{align}
\tilde{P}_{\psi_0} = \sqrt{1-\rho_0}\ket{\psi_0}\bra{\psi_0}+\hat{R},
\end{align}
where $\lVert \hat{R}\rVert=O\left(\epsilon'\right)$ ($\lVert \cdot \rVert$ stands for the spectral norm) for some value of $\epsilon'$ and we also have that $\hat{R}\ket{\psi_0} = 0$, thus,
\begin{align}
\left(\tilde{P}_{\psi_0}\right)^r = \left(1-\rho_0\right)^{r/2}\ket{\psi_0}\bra{\psi_0}+\hat{R}^r.
\end{align}
%
%
Therefore,
\begin{align}
\lVert \hat{R}^r \rVert = O(\epsilon'^r).
\end{align}
Here, we have illustrated that by applying this projector $r$ times we obtain a smaller projector residue that decreases exponentially with $r$.
This would also mean that the error on the state prepared decreases exponentially with respect to $r$.    

Now, we will estimate an upper bound on $r$. The state on the target register after $r$ iterations of the filtering operation from \Cref{eq:psi0_filter} is
\begin{align}
\ket{\psi} = \frac{\sum_i \phi_i \gamma^{r}_{i}\ket{\psi_i} }{\lVert\sum_i \phi_i \gamma^{r}_{i} \ket{\psi_i}\rVert}.
\end{align}
In \Cref{appendix:B3_derivation} we derive
\begin{align}\label{eq:epsilonT}
\epsilon = \Theta \left( \frac{\lVert\sum_{i\neq0} \phi_i \gamma^{r}_{i}\ket{\psi_i} \rVert }{\lVert\sum_i \phi_i \gamma^{r}_{i} \ket{\psi_i}\rVert}\right),
\end{align}
where
\begin{align}\label{eq:normalized_resulting}
 \left( \frac{\lVert\sum_{i\neq0} \phi_i \gamma^{r}_{i}\ket{\psi_i} \rVert }{\lVert\sum_i \phi_i \gamma^{r}_{i} \ket{\psi_i}\rVert}  \right)^2 &= \frac{\sum_{i\neq0} \lvert\phi_i\rvert^2 \lvert \gamma_{i}\rvert^{2r}}{\sum_{i} \lvert\phi_i\rvert^2 \lvert \gamma_{i}\rvert^{2r}}.
\end{align}
For $i\neq 0$, we have (see \Cref{appendix:useful_bounds} for bounds)
\begin{align}\label{eq:gamma_i_leq}
\lvert \gamma_{i}\rvert \leq \frac{1}{2^{m+1} \Delta}
\end{align}
where $\Delta$ is the lower bound on the spectral gap,
\begin{align}
\Delta_0 = \theta_1 - \theta_0 = \lambda E_1 - \lambda E_0.
\end{align}
Now, we derive an inequality for the quantity in \Cref{eq:normalized_resulting}:
\begin{align}\label{eq:eps_gamma0_inequality}
\frac{\sum_{i\neq0} \lvert\phi_i\rvert^2 \lvert \gamma_{i}\rvert^{2r}}{\sum_{i} \lvert\phi_i\rvert^2 \lvert \gamma_{i}\rvert^{2r}}&\leq \frac{1}{\lvert\phi_0\rvert^2\lvert\gamma_0\rvert^{2r}}\left(\frac{1}{2^{m+1}\Delta}\right)^{2r}.
\end{align}
In deriving this we have used \Cref{eq:gamma_i_leq}, $\sum_{i\neq0} \lvert\phi_i\rvert^2\leq 1$, and $\lvert\phi_i\rvert^2\geq 0$.
We would like to replace the term $\lvert\gamma_0\rvert^{2r}$ in \Cref{eq:eps_gamma0_inequality} with some more meaningful parameters like the probability of success,
\begin{align}\label{eq:prob_success}
P_r &= \lVert \tilde{P}^r_{\psi_0} \ket{\phi} \rVert^2.
\end{align}
As explained in \Cref{appendix:success_rate_rel}, $P_r$ approaches $\lvert \phi_0 \rvert^2$ from below as $\epsilon \to 0$. Thus, it is more convenient to parametrize $P_r$ the following way:
\begin{align}
P_r = (1-\rho) \lvert \phi_0 \rvert^2,
\end{align}
where $\rho \geq 0$. Also, in \Cref{appendix:success_rate_rel} we derive
\begin{align}\label{eq:overall_rate}
\lvert\gamma_{0}\rvert^{2r} = (1-\rho_0)^r = \Theta \left(1-\rho\right).
\end{align}
Thus, \Cref{eq:eps_gamma0_inequality} can be rewritten in terms of $\rho$ the following way
\begin{align}\label{eq:eps_rho_inequality}
\frac{\sum_{i\neq0} \lvert\phi_i\rvert^2 \lvert \gamma_{i}\rvert^{2r}}{\sum_{i} \lvert\phi_i\rvert^2 \lvert \gamma_{i}\rvert^{2r}}&\leq \frac{1}{\lvert\phi_0\rvert^2\Omega\left(1-\rho\right)}\left(\frac{1}{2^{m+1}\Delta}\right)^{2r}.
\end{align}
We can now relate $\epsilon$ to other relevant quantities through \Cref{eq:epsilonT} or more specifically
\begin{align}\label{eq:epsilonO}
 \epsilon^2 &= O\left(\frac{\sum_{i\neq0} \lvert\phi_i\rvert^2 \lvert \gamma_{i}\rvert^{2r}}{\sum_{i} \lvert\phi_i\rvert^2 \lvert \gamma_{i}\rvert^{2r}} \right).
\end{align}
Solving for $r$, we get
\begin{align}\label{eq:iterations_preliminary}
 r \leq \frac{\log O\left(\frac{1}{\epsilon \lvert\phi_0\rvert\sqrt{1-\rho}}\right)}{\log{2^{m+1}\Delta}}. 
\end{align}
Now, in order to ensure that the relative (to the overlap) error rate $\rho$ introduced by the $r$ filter operations is constant, we have to calculate the precision to which the ground energy has to be known.
Using \Cref{eq:overall_rate} and the bound $(1-\rho_0)^r\leq e^{-\rho_0 r}$ 
we can write
\begin{align}
\rho_0 \leq \frac{1}{r}\ln{\Omega\left(\frac{1}{1-\rho}\right)}.
\end{align}
Provided that $2\xi$ is less than the width of the main lobe from the base, we can bound $\lvert\gamma_{0}\rvert$ from above with a parabola. That is,
\begin{align}
\lvert \gamma_{0}\rvert^2 = (1-\rho_0) \leq 1-a 2^{2m}\xi^2.
\end{align}
Finally, this means that
\begin{align}
\xi &\leq \frac{1}{2^m}\sqrt{\frac{1}{a r}\ln{\Omega\left(\frac{1}{1-\rho}\right)}}\cr
&\leq \frac{1}{2^m}\sqrt{\frac{1}{a}\frac{\log{2^{m+1}\Delta}}{\log O \left(\frac{1}{\epsilon \lvert\phi_0\rvert\sqrt{1-\rho}}\right)} \ln{\Omega\left(\frac{1}{1-\rho}\right)}}, \cr
\end{align}
where $a$ has to be chosen such that the parabola has the same zero crossings as filter's main lobe. For the rectangular-window filter, $a=1$.

\onecolumngrid

\twocolumngrid

If we use the same method but applying the cosine-window filter, we run into a problem: $\rho_0$ is prohibitively large. That is, according to \Cref{eq:bound_rhor} the success rate becomes exponentially suppressed at $r \sim 5$ since the minimum $\rho_0$ possible is $\sim 0.2$. So, we propose a modification to the circuit in \Cref{fig:mqubit_pea} in order to create a coherent interference between the odd and even outcomes of the ancillary register. By applying a Hadamard gate on the least significant qubit, we create a constructive interference given the outcome of the least significant qubit is $0$ and a destructive one if the outcome is $1$. The resulting circuit can be seen in \Cref{fig:iteration_cosine}.
We can only do this coherent interference because we are using the centered version of the Fourier transform. Meaning that the amplitude distribution coming from the inverse Fourier transform interpolates smoothly between the discrete sample nodes; something we cannot say of the DFT with frequencies ranging $0,..,2^m-1$~\cite{ipol.2019.273}.

Finally, we can now turn again to the cosine-window case, but using the filter function $F_+(x)$ resulting from the circuit in \Cref{fig:iteration_cosine}; that is, we now have that
\begin{align}
\gamma_{i}&=F_{+}(2^m \theta^{(\xi)}_0-2^m \theta_i),
\end{align}
where
\begin{align}\label{eq:fplus}
F_{+}(q)&=\frac{F(q-1/2)+F(q+1/2)}{\sqrt{2}}.
\end{align}
An upper bound was derived for this function in \Cref{appendix:fxplus}, similar to the one in \Cref{eq:gamma_i_leq} for \Cref{eq:gx}. With this bound, we repeated the same procedure as above. Through this, we obtained analogous inequalities for the number of iterations:
\begin{align}
r_{\cos}\leq\frac{\log O\left(\frac{1}{\epsilon \lvert\phi_0\rvert\sqrt{1-\rho}}\right)}{\log{\frac{2^{3m+3}\Delta\left(\Delta+2^{-m}\right)\left(\Delta-2^{-m}\right)}{\pi^2}}},
\end{align}
and required precision:
\begin{widetext}
\begin{align}
\xi_{\cos} &\leq \frac{1}{2^m}\sqrt{\frac{1}{a_{\cos}}\frac{\log{\frac{2^{3m+3}\Delta\left(\Delta+2^{-m}\right)\left(\Delta-2^{-m}\right)}{\pi^2}}}{\log O\left(\frac{1}{\epsilon \lvert\phi_0\rvert\sqrt{1-\rho}}\right)} \ln{\Omega\left(\frac{1}{1-\rho}\right)}}.
\end{align}
\end{widetext}
Here, a similar bounding parabola was used, but with $a_{\cos}=1/4$.

It would appear that the required precision for the ground state energy value scales polylogarithmically with respect to all relevant quantities; however, we should notice the $1/2^m$ factor always appearing in front. For the case of the rectangular window, we have the condition $2^{m+1}\Delta>1$, which means that the required precision scales $\tilde{O}(\Delta)$. We have pointed out this fact in ~\Cref{tab:preparation_costs}, where the other methods show a similar scaling for the required ground state energy precision.

One way of estimating the ground state energy with the required precision is to perform the state preparation algorithm with an initial guess of the ground state energy $\theta^{(\xi)}_0=-0.5$. If not successful, we increase $\theta^{(\xi)}_0$ by the target precision $\xi=\tilde{O}(\Delta)$ and repeat the algorithm. The search can stop when one succeeds the state preparation algorithm at a rate that is not exponentially suppressed. This classical search method is something similar to what is proposed in Ref~\cite{sprep_Cirac} when the ground state energy is not known a priori.

So far in this section, we have shown that the iterative approach of state preparation through QPE has an exponential speed up over the single-iteration approach. As shown in \Cref{tab:preparation_costs}, the scaling with respect to the required precision, $\epsilon$, and the overlap, $\lvert\phi_0\rvert^{-1}$, has been improved exponentially.

Now, we will account for the error incurred by approximating the Hamiltonian evolution. Typically, the cost scaling of Hamiltonian evolution is quoted with respect to evolution operator error $\epsilon'$. The precise definition of evolution error is defined as
\begin{align}
\lVert e^{i2\pi\lambda H}-e^{i2\pi\lambda \tilde{H}} \rVert = \epsilon'.
\end{align}
We can now relate this evolution error to the Hamiltonian error
\begin{align}
\lVert H - \tilde{H} \rVert = O\left(\frac{\epsilon'}{2\pi\lambda}\right)
\end{align}
In order to properly estimate the cost with respect to the target error on the ground state, we turn to matrix perturbation theory in the Eigenvalue Problem. From Ref.~\cite{Wilkinson_EVP} we obtain the following bound
\begin{align}
\lVert|\tilde{\psi}_0\rangle - \ket{\psi_0} \rVert &\leq \frac{\lVert P_1 (\tilde{H}-H) \ket{\psi_0} \rVert}{E_1-E_0} \cr
&\leq\frac{\lambda\lVert \tilde{H}-H \rVert}{\Delta},
\end{align}
where $P_1$ is $\sum_{i\neq 0} \ket{\psi_i}\bra{\psi_i}$ and $|\tilde{\psi}_0\rangle$ is the ground state of the effective Hamiltonian $\tilde{H}$. Thus,
\begin{align}
\lVert|\tilde{\psi}_0\rangle - \ket{\psi_0} \rVert = O\left(\frac{\epsilon'}{\Delta}\right).
\end{align}
With this, we can now correctly estimate the cost of preparing the ground state. We have made a more precise treatment of the cost by looking how the ground state of the effective Hamiltonian perturbs the ground state. This introduces a modified cost factor $1/(\Delta \epsilon)$ as opposed to just $1/\epsilon$. This is an aspect that only until recently~\cite{YiCrosson_prod_pert} was taken into account for estimating ground state energy using QPEA and preparing states using digital adiabatic simulation.

All of this means that an algorithm which can simulate a Hamiltonian evolution $e^{i2\pi\lambda H}$ with an error $\epsilon'$ that costs $O(\text{polylog}(2^n,1/\epsilon'))$ will be sufficient to obtain a polylogarithmic cost in $1/(\Delta\epsilon)$ with our state preparation method. Any algorithm in Refs~\cite{Berry_2015_v1,Berry_2015_v2,Low_2017,Low2019hamiltonian} fulfills these requirements. The implied costs of using these algorithms for state preparation using the algorithms presented here are shown on the second row of \Cref{tab:preparation_costs}.

Generally speaking, methods based on product formulas alone do not meet the requirements. For example, in Ref~\cite{Berry_2006} authors used the $kth$-order product formulas provided by Suzuki in Ref~\cite{Suzuki1991GeneralTO}. The authors find that, in general, the number of product terms in the product expansion is $N_{exp}=O\left(5^{2k}/(\epsilon')^{1/2k}\right)$. From this formula, it is evident that we cannot increase $k$ arbitrarily as the cost increases exponentially with respect to $k$. The scaling of the cost with respect to $\epsilon$ and $\Delta$, using the Suzuki formulas, is reflected on the first row of \Cref{tab:preparation_costs}.

With the advent of optimal interpolation methods with the use of product formulas in Ref.~\cite{Low_Interp}, product formulas become another viable alternative even at asymptotically small error. We have also added the interpolation method on the second row of \Cref{tab:preparation_costs} as an alternative to LCU and Qubitization at similar overheads and cost.

\begin{table*}[ht]
 \centering
 \begin{tabularx}{\textwidth}{X|c|X|c}
 Preparation (ground energy known) & Gates & Qubits& Required precision  \\
 \hline
 This paper + Product formulas~\cite{Suzuki1991GeneralTO,Berry_2006} & $\begin{aligned}[t]\displaystyle \tilde O\left(\frac1{|\phi_0|^2\Delta^{1+\sfrac{1}{2k}}\epsilon^{\sfrac{1}{2k}}}\right)&\end{aligned}$ & {$\displaystyle O\left(\log N+\log\frac{1}{\Delta} \right)$}& $\tilde O(\Delta)$ \\
 This paper + LCU/Qubitization/Interp.~\cite{Berry_2015_v1,Berry_2015_v2,Low_2017,Low2019hamiltonian,Low_Interp}
 & $\displaystyle \tilde O\left(\frac1{|\phi_0|^2\Delta}\right)$ & 
 $\begin{aligned}\displaystyle O\left(\log N+\log\log\frac1{\Delta\epsilon}\quad\right.&\\\left.+\log\frac1\Delta \right)&\end{aligned}$& 
  $\tilde O(\Delta)$ \\
 Ge et al.~\cite{sprep_Cirac} & $\displaystyle \tilde O\left(\frac1{|\phi_0|^2\Delta}\right)$ & $\begin{aligned}\displaystyle O\left(\log N+\log\log\frac1{\Delta\epsilon}\quad\right.&\\\left.+\log\frac1\Delta \right)&\end{aligned}$ & $\tilde O(\Delta)$ \\
 Single-round Phase estimation + amp. amplif.~\cite{sprep_Cirac} & $\displaystyle \tilde O\left(\frac1{|\phi_0|^3\Delta\epsilon}\right)$ & $\begin{aligned}\displaystyle O\left(\log N+\log \frac1{\epsilon}\quad\right.&\\\left.+\log\frac1\Delta\right)&\end{aligned}$ & $O\left(|\phi_0|\epsilon\Delta\right)$\\ 
\end{tabularx}
\caption{Algorithms for ground state preparation for the case when the ground energy is known beforehand to the required precision. For simplicity of comparison, we have omitted speed-ups in $|\phi_0|$ through amplitude amplification. We have also omitted some overhead costs, like for example the base oracle gate cost in the oracle based approach of Hamiltonian simulation~\cite{Berry_2015_v1,Berry_2015_v2}.} \label{tab:preparation_costs}
\end{table*}

\section{Numerical Tests of State Preparation\label{sec:results_sprep}}

The theory that will serve as a test ground will be the $(1+1)$-dimensional massive Thirring model

\begin{align} 
\label{eq-action-thirring}
    S_{\mathrm{Th}}[\psi,\bar{\psi}] \
    = \int d^2x &\left[ \bar{\psi}i \gamma^{\mu}\partial_{\mu}\psi \
      - m\,\bar{\psi}\psi \right. \cr
      &\left.-\frac{g}{2} \left( \bar{\psi}\gamma_{\mu}\psi \right)\left( \bar{\psi}\gamma^{\mu}\psi \right) \right] \,,
\end{align}

For our simulations, we use the lattice Hamiltonian derived in \cite{Ba_uls_2019}, 

\begin{align}
\label{eq:H_sim}
 \bar{H}_{{\mathrm{sim}}} &= -\frac{1}{2} \sum_{n}^{N-2} \left( S_{n}^{+}S_{n+1}^{-} 
            + S_{n+1}^{+}S_{n}^{-} \right) \cr
            &+ a \tilde{m}_{0} \sum_{n}^{N-1} \left(-1\right)^{n} \left(
              S_{n}^{z}+\frac{1}{2} \right) \cr 
        &+ \Delta (g) \sum_{n}^{N-1} \left( S_{n}^{z}+\frac{1}{2} \right) \ 
            \left( S_{n+1}^{z}+\frac{1}{2} \right).
\end{align}

In order to boost the overlap of our initial guess with the ground state we use a variational approach. The ansatz for the ground state consists of alternating non-commuting operators that comprise the Hamiltonian~\cite{farhi2014quantum,Wecker2015,farhi2019quantum,Zhou2019,Ho_2019}:
\begin{align}
| \phi (\bm{\alpha},\bm{\beta},\bm{\gamma})  \rangle &= e^{-i \gamma_p H_{\gamma}} e^{- i \beta_p H_{\beta}} e^{-i \alpha_p H_{\alpha}} \cdots \cr
&e^{-i \gamma_1 H_{\gamma}} e^{- i \beta_1 H_{\beta}} e^{-i \alpha_1 H_{\alpha}} |\phi_0 \rangle,
\label{eqn:QAOA}
\end{align}
where
\begin{align}
H_{\alpha} &=-\frac{1}{2} \sum_{n=0} \left( S_{2n}^{+}S_{2n+1}^{-} + S_{2n+1}^{+}S_{2n}^{-} \right) \\
H_{\beta} &=-\frac{1}{2} \sum_{n=1} \left( S_{2n-1}^{+}S_{2n}^{-} + S_{2n}^{+}S_{2n-1}^{-} \right) \cr
H_{\gamma} &= a \tilde{m}_{0} \sum_{n}^{N-1} \left(-1\right)^{n} \left( S_{n}^{z}+\frac{1}{2} \right) \cr 
            &+ \Delta (g) \sum_{n}^{N-1} \left( S_{n}^{z}+\frac{1}{2} \right) \left( S_{n+1}^{z}+\frac{1}{2} \right).            
\end{align}

We proceed to minimize

\begin{align}
\bra{\phi (\bm{\alpha},\bm{\beta},\bm{\gamma})} \bar{H}_{{\mathrm{sim}}} \ket{\phi (\bm{\alpha},\bm{\beta},\bm{\gamma})}
\end{align}

With these new $\ket{\phi (\bm{\alpha},\bm{\beta},\bm{\gamma})}$ states we then follow the state preparation procedure in the last section for both the cosine window and rectangular window.

We would like to corroborate the bounds in the previous section. We would like to see two things: an exponential decay in the error with respect to the number of iterations, and an improvement by using the cosine window.

We also wish to separate the error from the implementation of the Hamiltonian evolution and the error from the projection procedure. For that reason, we perform the projection procedure with respect to the effective Hamiltonian resulting from the product expansion (summation series if we were using Qubitization, for example). That is,
\begin{align}
U=e^{i2\pi\lambda H_{eff}}=U^{d}_{step},
\end{align}
or
\begin{align}\label{eq:Heff}
H_{eff} = -i\frac{d}{2\pi\lambda}\log { U_{step} },
\end{align}
where
\begin{align}
U_{step} = U^{i\frac{\Delta t}{2} H_\alpha} U^{i\frac{\Delta t}{2} H_\beta} U^{i\Delta t H_\gamma} U^{i\frac{\Delta t}{2} H_\beta} U^{i\frac{\Delta t}{2} H_\alpha}.
\end{align}

This last equation is the second-order Suzuki's product expansion~\cite{Suzuki1991GeneralTO} for \Cref{eq:H_sim}. Here, $\Delta t=2\pi\lambda/d$ and $d$ is an integer equal or greater to $1$.

\begin{figure}[!htb]
\centering
\includegraphics[width=0.3\textwidth]{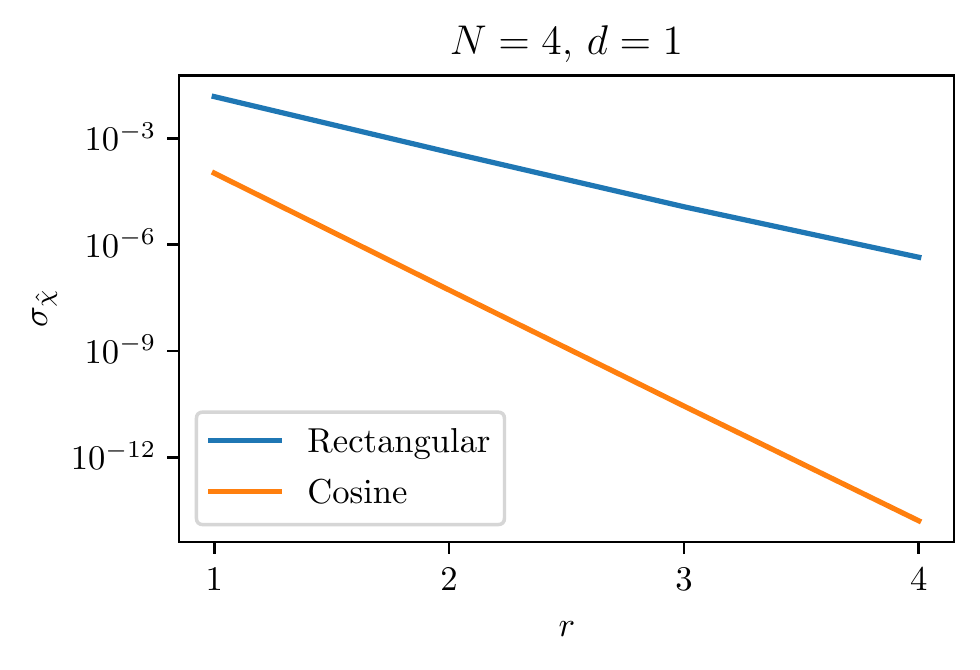} \includegraphics[width=0.3\textwidth]{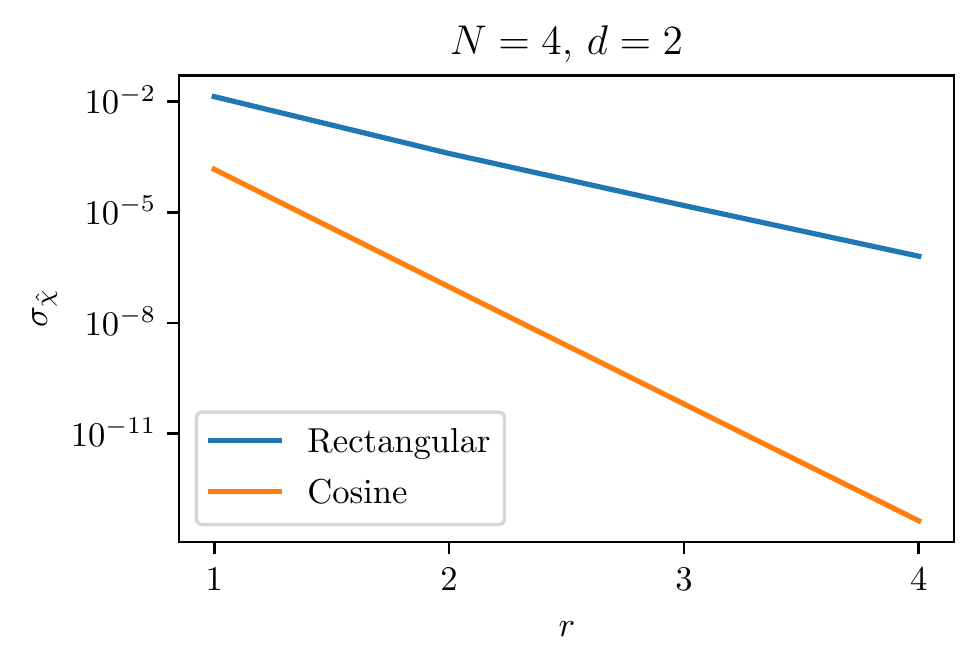} \includegraphics[width=0.3\textwidth]{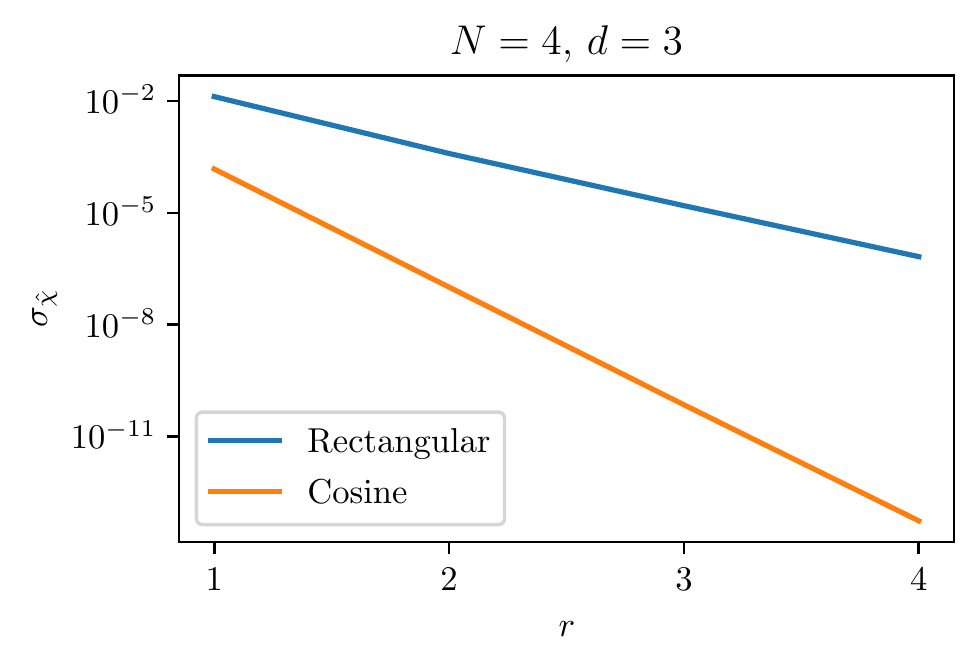} \\
\includegraphics[width=0.3\textwidth]{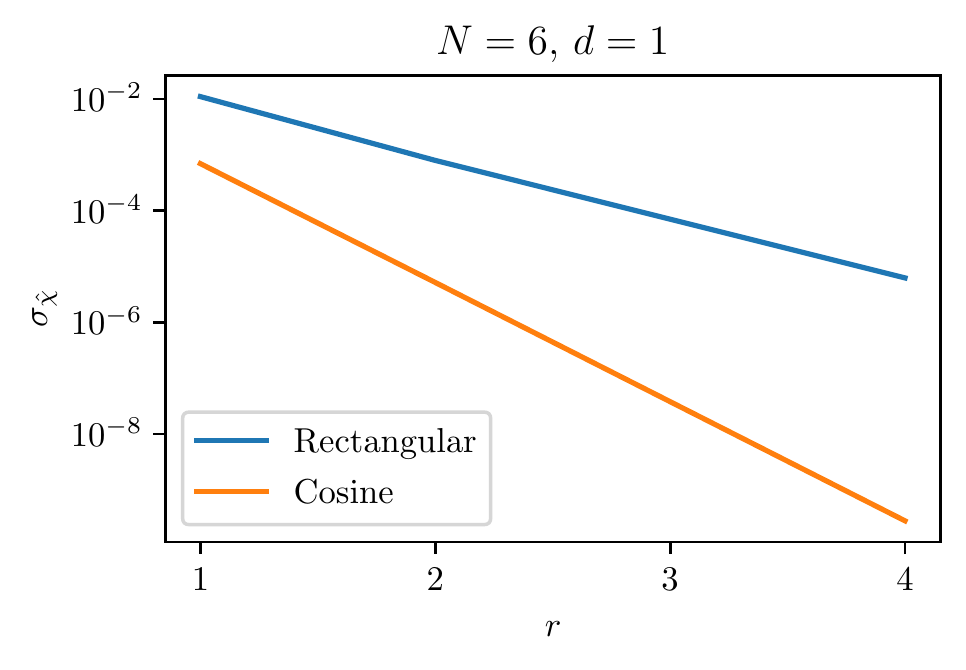} \includegraphics[width=0.3\textwidth]{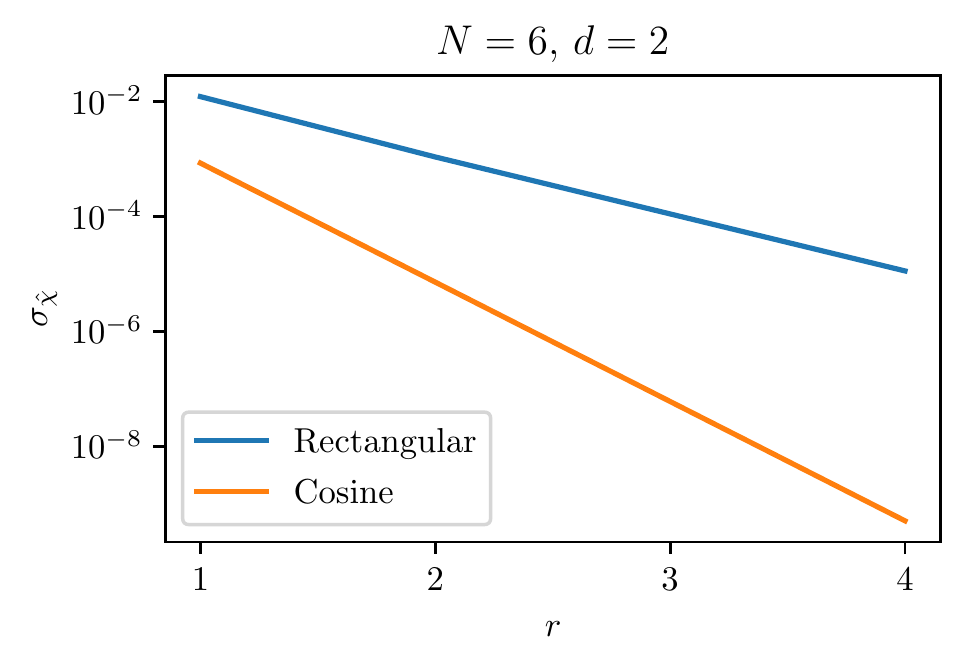} \includegraphics[width=0.3\textwidth]{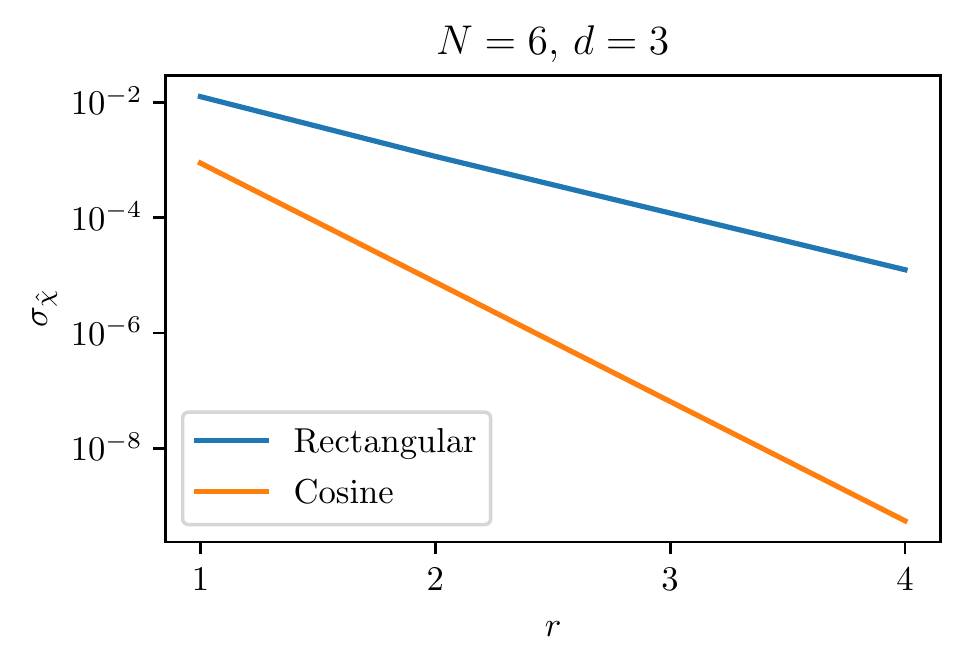} \\
\includegraphics[width=0.3\textwidth]{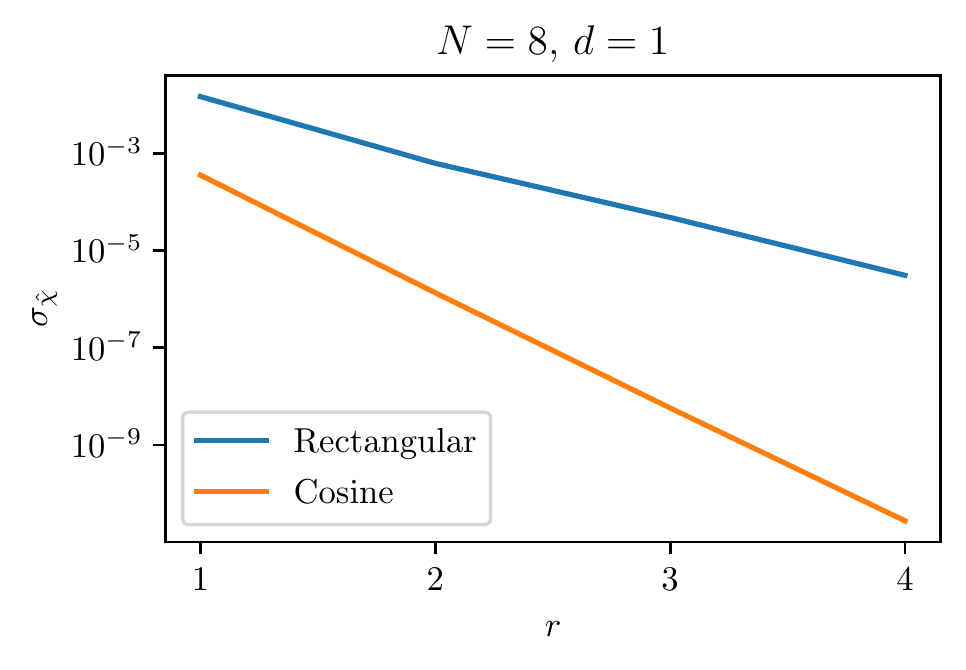} \includegraphics[width=0.3\textwidth]{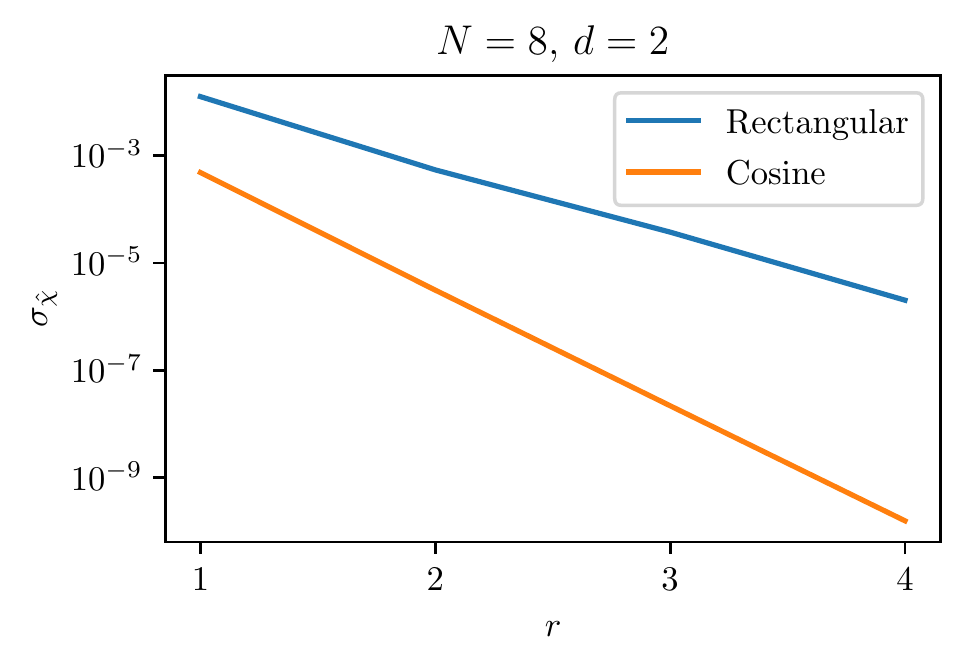} \includegraphics[width=0.3\textwidth]{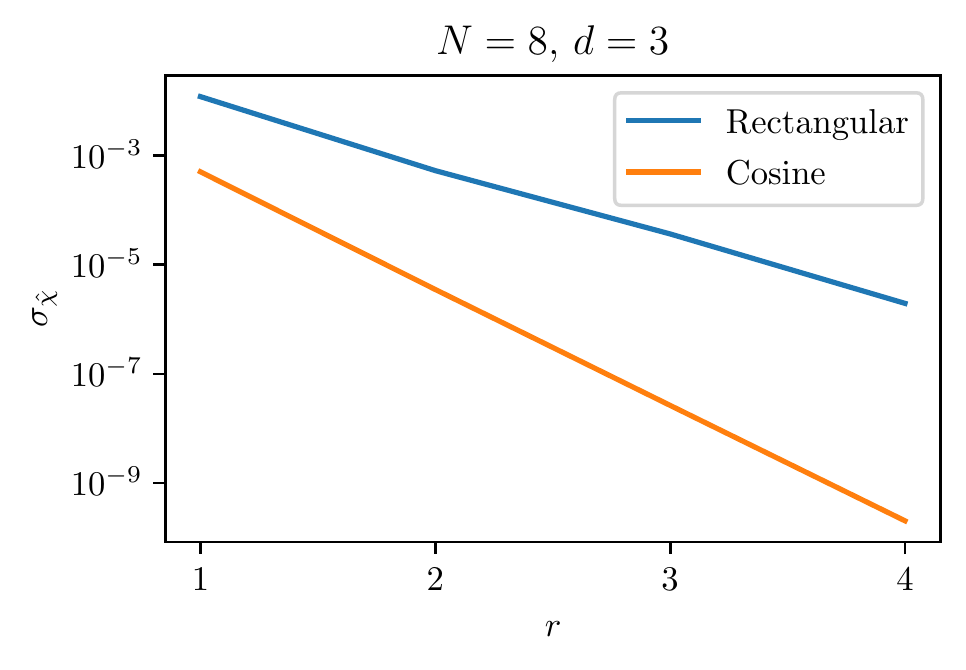} \\
\caption{Here we show comparisons of the excited state contamination, $\sigma_{\chi}$, in the chiral condensate for the values $d\in{1,2,3}$ and $N\in{4,6,8}$. The excited state contamination is plotted against the number of iterations, $r$, showing that the cosine-window filter achieves an improvement over the rectangular-window one.\label{fig:sigma_chi}}
\end{figure}

In order to estimate the error $\epsilon$ we evolve $\ket{\lambda}$ and measure the expectation value of an operator of interest $\hat{O}$. That is,
\begin{align}
\mathbf{O} = \left\{\bra{\lambda}(U_{step}^\dag)^n\hat{O}U_{step}^n\ket{\lambda}:n=\{1,2,3\dots,N\}\right\}
\end{align}
We then estimate a central value defined as the mean:
\begin{align}
\bar{O}=\mathbf{E}\left[\mathbf{O}\right]
\end{align}
and a standard deviation
\begin{align}
\sigma_O=\sqrt{\mathbf{E}\left[\left(\mathbf{O}-\bar{O}\right)^2\right]}
\end{align}
For this numerical test we have chosen the chiral condensate as our observable.
\begin{align}
\hat{\chi}=\frac{1}{n}\sum_{i=0}(-1)^{i+1}Z_i
\end{align}
In \Cref{fig:sigma_chi} we show a comparisons of $\sigma_{\chi}$ versus the number of iterations, $r$, for both the rectangular and the cosine window. This comparison is done for $d\in \{ 1,2,3 \}$ and system sizes $N\in \{ 4,6,8 \}$. We would also like to stress that $\sigma_{\chi}$ estimates the excited state contamination with respect to the effective Hamiltonian in \Cref{eq:Heff}. The two main takeaways of these numerical tests are that: $\sigma_{\chi}$ appears to decay exponentially with respect to the number of iterations, and that the cosine window outperforms the rectangular window. \\

\clearpage

\onecolumngrid

\twocolumngrid

\section{Conclusion}

In this work, we have presented the effects of using a cosine tapering window on the quantum phase estimation algorithm. It was demonstrated that one obtains a cubic improvement in gate complexity scaling with respect to the error rate. It is left to future research the exploration of other windows, their optimization with respect to other metrics, as well as the exploration of possible hybrid quantum-classical approaches. 

We also showed the effects of this window when the phase estimation algorithm is re-purposed for quantum state preparation. Simultaneously, we showed that using repeated blunted filter operations was more efficient than performing a single sharper filter operation. The improvements were exponential in $\lvert\phi_0\rvert^{-1}$ and $\epsilon^{-1}$. Nevertheless, the linear scaling with respect to $\Delta^{-1}$ remains. 

\section{Acknowledgements}

The authors would like to thank Ning Bao, Andreas Hackl, Daniel Kn\"uttel, Christoph Lehner, Akio Tomiya, and Nathan Wiebe for the useful discussions. 
During this work, G.~R. and T.~I. were supported through Brookhaven National Laboratory, the Laboratory Directed Research and Development (LDRD) program No. 21-043, 
by Program Development Fund No. NPP PD 19-025,
and by the U.S. Department of Energy, Office of Science, Office of Nuclear Physics, also under the Contract No. DE-SC0012704 (BNL).
Y.~K. is supported by the U.S. Department of Energy, Office of Science, National Quantum Information Science Research Centers, Co-design Center for Quantum Advantage, under the contract DE-SC0012704 (BNL).


\appendix

\section{Detailed derivations of some formulas}

\subsection{$F_{+}(x)$ and its upper bound\label{appendix:fxplus}}

We convert the filter function, $F_+(q)$~\eqref{eq:fplus}, to upper-bound its magnitude.

\begin{align}
&F_{+}(q)=\frac{F(q-1/2)+f(q+1/2)}{\sqrt{2}}\cr
&=\frac{\sin \left(\frac{\pi }{2^m}\right) \cos (\pi  q - \frac{\pi}{2}) }{2^{m+1} \sin \left(\frac{2\pi -2 \pi  q}{2^{m+1}}\right) \sin \left(\frac{2 \pi  q }{2^{m+1}}\right)} \cr
&+ \frac{\sin \left(\frac{\pi }{2^m}\right) \cos (\pi  q + \frac{\pi}{2}) }{2^{m+1} \sin \left(\frac{-2 \pi  q}{2^{m+1}}\right) \sin \left(\frac{2\pi+2 \pi  q }{2^{m+1}}\right)} \cr
&=\frac{\sin \left(\frac{\pi }{2^m}\right) \sin (\pi  q) \left( \sin \left(\frac{\pi + \pi  q}{2^{m}}\right) + \sin \left(\frac{\pi - \pi  q}{2^{m}}\right) \right)}{2^{m+1} \sin \left(\frac{\pi  q }{2^{m}}\right) \sin \left(\frac{\pi - \pi  q}{2^{m}}\right) \sin \left(\frac{\pi + \pi  q}{2^{m}}\right) } \cr
&=\frac{\sin \left(\frac{\pi }{2^m}\right) \sin (\pi  q)\sin \left(\frac{\pi}{2^m}\right)\cos\left(\frac{ \pi  q}{2^{m}}\right)}{2^{m} \sin \left(\frac{\pi  q }{2^{m}}\right) \sin \left(\frac{\pi - \pi  q}{2^{m}}\right) \sin \left(\frac{\pi + \pi  q}{2^{m}}\right) } \cr
&=\frac{\sin^2 \left(\frac{\pi }{2^m}\right) \sin (\pi  q)\cot\left(\frac{ \pi  q}{2^{m}}\right)}{2^{m} \sin \left(\frac{\pi - \pi  q}{2^{m}}\right) \sin \left(\frac{\pi + \pi  q}{2^{m}}\right) }.
\end{align}
Hence, the absolute value of $F_+(q)$ is upper-bounded as,
\begin{align}\label{eq:bound_fxplus}
&\left\vert F_{+}(q) \right\vert 
\cr
&\leq \frac{\sin^2 \left(\frac{\pi }{2^m}\right)}{2^{m} \lvert\sin \left(\frac{\pi  q }{2^{m}}\right)\rvert \lvert\sin \left(\frac{\pi - \pi  q}{2^{m}}\right)\rvert \lvert\sin \left(\frac{\pi + \pi  q}{2^{m}}\right) \rvert } \cr
&\leq \frac{\left( 2^{m-1} \right)^3\sin^2 \left(\frac{\pi }{2^m}\right)}{2^{m} \lvert q\rvert \lvert q-1\rvert \lvert q+1\rvert }, \ \text{for}\  0 \leq 1+\lvert q\rvert \leq 2^{m-1} \cr
&\leq \frac{ \pi^2}{8 \lvert q\rvert \lvert q-1\rvert \lvert q+1\rvert }, \ \text{for}\  0 \leq 1+\lvert q\rvert \leq 2^{m-1}. \cr
\end{align}

\subsection{Some useful bounds\label{appendix:useful_bounds}}

Using the inequality,
\begin{align}
\big|\sin\frac{x}{2}\big|\geq\dfrac{|x|}{\pi} \ \text{for}\ 0\leq|x|\leq\pi,
\end{align}
we provide an upper bound on the Dirichlet kernel,
\begin{align}
D_{M}(x)=\frac{\sin \left(M \pi x \right)}{M\sin \left(\pi x\right)}.
\end{align}
Its magnitude bounded as
\begin{align}
|D_{M}(x)|=&\dfrac{1}{M}\dfrac{|\sin M \pi x|}{|\sin\pi x|}\cr
&\leq\dfrac{1}{M}\dfrac{1}{|\sin \pi x|}\leq\dfrac{1}{ 2 M |x|},
\end{align}
for $0\leq|x|\leq1/2$.

\subsection{Useful relations of success rate\label{appendix:success_rate_rel}}

The probability of successfully applying the filter in \Cref{eq:psi0_filter} $r$ times is
\begin{align}
P_r &= \lVert P^r_{\psi_0} \ket{\phi} \rVert^2 \cr 
&=\lVert\sum_{i} \phi_i \gamma^{r}_{i}\ket{\psi_i} \rVert^2 = \sum_{i} \lvert\phi_i\rvert^2 \lvert \gamma_{i}\rvert^{2r}
\cr
&= \lvert \gamma_{0}\rvert^{2r} \lvert\phi_0\rvert^2 + \sum_{i\neq0} \lvert\phi_i\rvert^2 \lvert \gamma_{i}\rvert^{2r} \cr
&= \lvert \gamma_{0}\rvert^{2r} \lvert\phi_0\rvert^2 + P_r \frac{\lVert\sum_{i\neq0} \phi_i \gamma^{r}_{i}\ket{\psi_i} \rVert^2 }{\lVert\sum_i \phi_i \gamma^{r}_{i} \ket{\psi_i}\rVert^2} \cr
&= \lvert \gamma_{0}\rvert^{2r} \lvert\phi_0\rvert^2 + P_r O(\epsilon^2)
\end{align}
To get to the last line we have used \Cref{eq:epsilon}. Solving for $P_r$, we obtain that
\begin{align}\label{eq:Pr_epsilon}
 P_r &= \frac{\lvert \gamma_{0}\rvert^{2r}}{1-O(\epsilon^2)} \lvert\phi_0\rvert^2 
\end{align}
It is clear that $P_r$ approaches $\lvert \phi_0 \rvert^2$ from below as $\epsilon \to 0$. Therefore, we find it convenient to parameterize $P_r$ the following way
\begin{align}
P_r = (1-\rho) \lvert \phi_0 \rvert^2,
\end{align}
where $\rho \geq 0$.
From this definition and \Cref{eq:Pr_epsilon}, it immediately follows that
\begin{align}
 \lvert\gamma_{0}\rvert^{2r} &\leq \left(1-\rho\right)
\end{align}
More generally, 
\begin{align}\label{eq:gamma_vs_rho}
\lvert\gamma_{0}\rvert^{2r} \to \frac{P_r}{\lvert\phi_0\rvert^2}, \text{ as } \epsilon\to 0. 
\end{align}
Equivalently, using definition in \Cref{eq:Pr_epsilon},
\begin{align}\label{eq:gamma_vs_rho_theta}
\lvert\gamma_{0}\rvert^{2r} = \Theta\left(1-\rho\right).
\end{align}

\subsection{Asymptotic expression for $\epsilon$}
\label{appendix:B3_derivation}
First, we assume that we can choose $\ket{\psi_0}$ such that the product $\gamma_0^r\phi_0$ is always real and positive. Thus, we can obtain the following bound for $\epsilon$

\begin{align}\label{eq:epsilon}
\epsilon&\equiv \lVert \ket{\psi} - \ket{\psi_0} \rVert \cr 
&= \left\lVert \frac{\sum_i \phi_i \gamma^{r}_{i} \ket{\psi_i}}{\lVert\sum_i \phi_i \gamma^{r}_{i} \ket{\psi_i}\rVert} - \ket{\psi_0} \right\rVert \cr
&=\frac{\lVert(\gamma^r_0\phi_0-\sqrt{P_r})\ket{\psi_0}+\sum_{i\neq0}\phi_i \gamma^{r}_{i} \ket{\psi_i}\rVert}{\lVert\sum_i \phi_i \gamma^{r}_{i} \ket{\psi_i}\rVert} \cr
&\geq  \frac{\lVert\sum_{i\neq0} \phi_i \gamma^{r}_{i}\ket{\psi_i} \rVert }{\lVert\sum_i \phi_i \gamma^{r}_{i} \ket{\psi_i}\rVert}.
\end{align}

Now, if we use \Cref{eq:gamma_vs_rho_theta} we also obtain that

\begin{align}
\lVert \ket{\psi} - \ket{\psi_0} \rVert &= \Theta \left(\frac{\lVert\sum_{i\neq0} \phi_i \gamma^{r}_{i}\ket{\psi_i} \rVert }{\lVert\sum_i \phi_i \gamma^{r}_{i} \ket{\psi_i}\rVert}\right)
\end{align}

\subsection{Change of bases in $QFT$}
\label{appendix:change_of_bases}
\begin{widetext}
\begin{align}
QFT&=\frac{1}{\sqrt{2^m}}\sum^{2^{m}-1}_{n=0} \left( \sum^{2^{m-1}-1}_{k=0} e^{-\frac{2 \pi i n k}{2^m}} \ket{k}\bra{n}+ \sum^{2^{m}-1}_{k=2^{m-1}} e^{-\frac{2 \pi i n k}{2^m}} \ket{k}\bra{n} \right) \cr
&=\frac{1}{\sqrt{2^m}}\sum^{2^{m}-1}_{n=0} \left( \sum^{2^{m-1}-1}_{k=0} e^{-\frac{2 \pi i n k}{2^m}} \ket{k}\bra{n}+ \sum^{-1}_{q=-2^{m-1}} e^{-\frac{2 \pi i n (2^m+q)}{2^m}} \ket{n}\bra{q} \right) \cr
&=\frac{1}{\sqrt{2^m}}\left( \sum^{2^{m-1}-1}_{n=0}  \sum^{2^{m-1}-1}_{k=0} e^{-\frac{2 \pi i n k}{2^m}} \ket{k}\bra{n}+ \sum^{2^{m-1}-1}_{n=0} \sum^{-1}_{q=-2^{m-1}} e^{-\frac{2 \pi i n q}{2^m}} \ket{n}\bra{q} \right.\cr
&\left.+\sum^{2^{m}-1}_{n=2^{m-1}}  \sum^{2^{m-1}-1}_{k=0} e^{-\frac{2 \pi i n k}{2^m}} \ket{k}\bra{n}+ \sum^{2^{m}-1}_{n=2^{m-1}} \sum^{-1}_{q=-2^{m-1}} e^{-\frac{2 \pi i n q}{2^m}} \ket{n}\bra{q} \right) \cr
&=\frac{1}{\sqrt{2^m}}\left( \sum^{2^{m-1}-1}_{n=0}  \sum^{2^{m-1}-1}_{k=0} e^{-\frac{2 \pi i n k}{2^m}} \ket{k}\bra{n}+ \sum^{2^{m-1}-1}_{n=0} \sum^{-1}_{q=-2^{m-1}} e^{-\frac{2 \pi i n q}{2^m}} \ket{n}\bra{q} \right.\cr
&\left.+\sum^{-1}_{x=-2^{m-1}}  \sum^{2^{m-1}-1}_{k=0} e^{-\frac{2 \pi i (2^m+x) k}{2^m}} \ket{x}\bra{k}+ \sum^{-1}_{x=-2^{m-1}} \sum^{-1}_{q=-2^{m-1}} e^{-\frac{2 \pi i (2^m+x) q}{2^m}} \ket{q}\bra{x} \right) \cr
&= \frac{1}{\sqrt{2^m}} \sum^{2^{m-1}-1}_{x=-2^{m-1}} \sum^{2^{m-1}-1}_{q=-2^{m-1}} e^{-\frac{2\pi i x q}{2^m}}\ket{q}\bra{x}
\end{align}
\end{widetext}

\subsection{Inequalities for Phase Estimation\label{appendix:qubit_bound}}

Here, we find an upper bound on the amplitude $F(q)$~\Cref{Fq}:
\begin{widetext}
\vspace{4em}
\begin{align}\label{eq:bound_fx}
\left\vert F(q;m) \right\vert &\leq \frac{\sin \left(\frac{\pi }{2^m}\right)}{\sqrt{2}\lvert\sin \left(\frac{\pi+2 \pi  q }{2^{m+1}}\right)\rvert} \dfrac{1}{2 \lvert 1/2 -q \rvert }\ \text{for}\  0 \leq \lvert1/2-q\rvert \leq 2^{m-1} \cr
&\leq \frac{\sin \left(\frac{\pi }{2^m}\right)}{\sqrt{2} \lvert\frac{1+2   q }{2^{m}}\rvert} \dfrac{1}{ 2 |1/2-q|} \ \text{for}\  0 \leq \lvert1/2-q\rvert \leq 2^{m-1}, \ \text{and for}\  0 \leq \lvert1/2+q\rvert \leq 2^{m-1} \cr
&\leq \frac{\sin \left(\frac{\pi }{2^m}\right) 2^m}{\sqrt{2} 4 } \dfrac{1}{ \lvert 1/2 + q \rvert \lvert 1/2-q\rvert} \ \text{for}\  0 \leq \lvert q\rvert + 1/2 \leq 2^{m-1}
\end{align}
\end{widetext}
where we have used the fact that $\lvert \sin\left(x\right)\rvert$ is $\pi$-periodic in order to get a similar inequalities when $\lvert x\rvert + 1/2 > 2^{m-1}$.

Therefore, the probability of getting an error greater than $\frac{k}{2^m}$ is
\begin{widetext}
\begin{align}
e&=\sum_{k\leq l < 2^{m-1} } \lvert \alpha_{lz} \rvert^2 + \sum_{-2^{m-1}\leq l < k } \lvert \alpha_{lz} \rvert^2 \leq c^2\sum_{k\leq l < 2^{m-1} } \left( \frac{1}{(l-\delta 2^m+1/2)(l-\delta 2^t-1/2)} \right)^2 \cr
&+ c^2\sum_{-2^{m-1}< l < -k } \left( \frac{1}{(l-\delta 2^m+1/2)(l-\delta 2^m-1/2)} \right)^2 +c^2\left(\frac{1}{\lvert-2^{m-1}-\delta 2^m+1/2\rvert\lvert2^{m-1}+\delta 2^m+1/2)\rvert}\right)^2 \cr
&= c^2\sum^{2^{m-1}-2}_{l=k-1} \left( \frac{1}{(l-\delta 2^m+1/2+1)(l-\delta 2^m-1/2+1)} \right)^2 \cr
&+ c^2\sum^{2^{m-1}-2}_{l=k-2} \left( \frac{1}{(l+\delta 2^m-1/2+1)(l+\delta 2^m+1/2+1)} \right)^2 +c^2\left(\frac{1}{\lvert2^{m-1}+\delta 2^m-1/2\rvert\lvert2^{m-1}+\delta 2^m+1/2)\rvert}\right)^2 \cr
&\leq c^2\sum^{2^{m-1}-2}_{l=k-1} \frac{1}{l^4} 
+ c^2\sum^{2^{m-1}-1}_{l=k-2} \frac{1}{l^4} 
\leq 2c^2\sum^{2^{m-1}-1}_{l=k-1} \frac{1}{l^4} 
\leq 2c^2\int^{2^{m-1}-1}_{l=k-2} \frac{dl}{l^4} \cr
&= \frac{2c^2}{3(k-2)^3} < \frac{\pi^2}{48 (k-2)^3}.
\end{align}%
\end{widetext}

For the term in the fourth line, we have used the fact that $\lvert \sin \left( x \right) \rvert$ is $\pi$-periodic to find a bound when $l=-2^{m}$. 
Therefore, to get an estimate that is within $k/2^{m}=2^{p-1}/2^{m}=1/2^{t+1}$ of the value of $\theta_i$ with an error rate of at most $e$, we need the total number qubits $m$ to be
\begin{align}
m=t+p=t+\left\lceil\log_2\left(\frac{\pi^{2/3}}{48^{1/3} e^{1/3}}+2\right)\right\rceil.
\end{align}

\onecolumngrid

\section{Ancillary Circuits}

\begin{figure}[!htb]
\scalebox{0.8}{
\begin{quantikz}
\lstick[wires=5]{$\ket{0}^{\tp m}$} & \qw & \gate[wires=5,nwires=3]{QFT^{-1}} & \qw
 \\
  & \qw &\qw & \qw
 \\
  & \vdots & \vdots &
 \\
  & \qw &\qw & \qw
 \\
  & \qw & \qw& \qw
\end{quantikz}
}
\(\rightarrow\)
\centering
\scalebox{0.8}{
  \begin{quantikz}
  \lstick[wires=5]{$\ket{0}^{\tp m}$}& \gate{H} & \qw
  \\
  & \gate{H} & \qw
  \\
  & \vdots
  \\
  & \gate{H} & \qw
  \\  
  & \gate{H} & \qw 
  \\
  \end{quantikz}
}
\caption{Here, we have taken the circuit up to the black dashed line in \Cref{fig:mqubit_pea_textbook} and simplified it given the initial state; same goes for \Cref{fig:iteration_rectangular}. }\label{fig:simp_rectangular}
\end{figure}
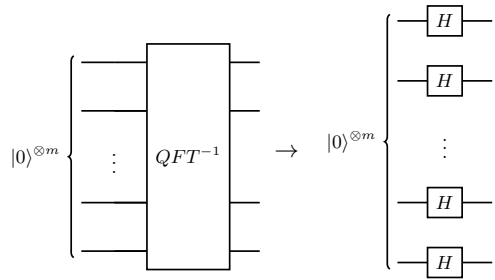

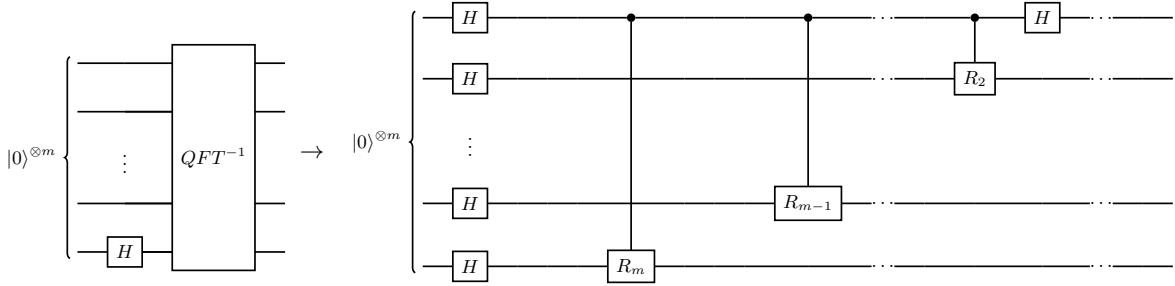
\begin{figure}[!htb]
\scalebox{0.8}{
\begin{quantikz}
\lstick[wires=5]{$\ket{0}^{\tp m}$} & \qw & \gate[wires=5,nwires=3]{QFT^{-1}} & \qw
 \\
  & \qw &\qw & \qw
 \\
  & \vdots & \vdots &
 \\
  & \qw &\qw & \qw
 \\
  & \gate{H}& \qw& \qw
\end{quantikz}
}
\(\rightarrow\)
\centering
\scalebox{0.8}{
  \begin{quantikz}
  \lstick[wires=5]{$\ket{0}^{\tp m}$}& \gate{H} & \qw & \qw & \qw & \ctrl{4} & \qw & \qw & \qw & \ctrl{3} & \qw\ldots & \qw & \ctrl{1} & \gate{H} & \qw\ldots & \qw & \qw &
  \\
  & \gate{H} & \qw & \qw & \qw & \qw & \qw & \qw & \qw & \qw & \qw\ldots & \qw & \gate{R_{2}} &  \qw  & \qw\ldots & \qw & \qw &
  \\
  & \vdots
  \\
  & \gate{H} & \qw & \qw & \qw & \qw & \qw & \qw & \qw & \gate{R_{m-1}} & \qw\ldots & \qw & \qw & \qw & \qw\ldots & \qw & \qw &
  \\  
  & \gate{H} & \qw & \qw & \qw & \gate{R_{m}} & \qw & \qw & \qw & \qw & \qw\ldots & \qw & \qw & \qw & \qw\ldots & \qw & \qw &
  \\
  \end{quantikz}
}
\caption{Here, we have taken the circuit up to the black dashed line in \Cref{fig:mqubit_pea} and simplified it given the initial state; same goes for \Cref{fig:iteration_cosine}.}\label{fig:simp_cosine}
\end{figure}

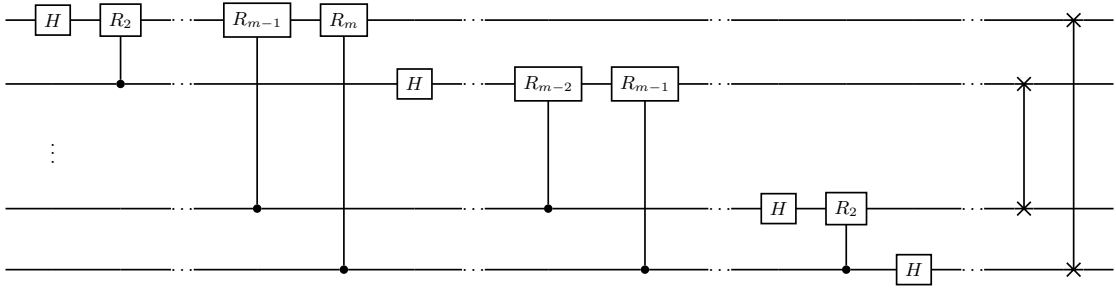
\begin{figure}[!htb]
\centering
\scalebox{0.8}{
  \begin{quantikz}
  & \gate{H} & \gate{R_2} & \qw\ldots & \gate{R_{m-1}} & \gate{R_{m}} & \qw & \qw\ldots & \qw & \qw & \qw\ldots & \qw & \qw & \qw & \qw\ldots & \qw & \swap{4} & \qw  \\
  & \qw & \ctrl{-1} & \qw\ldots & \qw & \qw & \gate{H} & \qw\ldots & \gate{R_{m-2}} & \gate{R_{m-1}} & \qw\ldots & \qw & \qw & \qw & \qw\ldots & \swap{2} & \qw & \qw \\
  & \vdots \\
  & \qw & \qw & \qw\ldots & \ctrl{-3} & \qw & \qw & \qw\ldots & \ctrl{-2} & \qw & \qw\ldots & \gate{H} & \gate{R_{2}} &  \qw  & \qw\ldots & \targX{} & \qw & \qw \\
  & \qw & \qw & \qw\ldots & \qw & \ctrl{-4} & \qw & \qw\ldots & \qw & \ctrl{-3} & \qw\ldots & \qw & \ctrl{-1} & \gate{H} & \qw\ldots & \qw & \targX{} & \qw
  \end{quantikz}
}
\caption{Our convention for $QFT^{-1}$. The corresponding $QFT$ can be obtained by simply inverting the circuit. This circuit is most widely labeled as $QFT$ due to conventions in quantum computing literature~\cite{nielsen2002quantum,Cleve_1998}.\label{fig:qft_convention}}
\end{figure}

\Cref{fig:simp_rectangular,fig:simp_cosine} provide ancillary circuits for the preparation of the filter distributions. \Cref{fig:qft_convention} just establishes the convention for $QFT$.



\twocolumngrid

\bibliographystyle{utphys-noitalics}

\begin{thebibliography}{10}

\bibitem{Abrams_QPE}
D.~S. {Abrams} and S.~{Lloyd}, ``{Quantum Algorithm Providing Exponential Speed
  Increase for Finding Eigenvalues and Eigenvectors},''
  \href{http://dx.doi.org/10.1103/PhysRevLett.83.5162}{\prl {\bfseries 83}
  no.~24, (Dec., 1999) 5162--5165},
  \href{http://arxiv.org/abs/quant-ph/9807070}{{\ttfamily
  arXiv:quant-ph/9807070 [quant-ph]}}.

\bibitem{Cleve_1998}
R.~Cleve, A.~Ekert, C.~Macchiavello, and M.~Mosca, ``Quantum algorithms
  revisited,'' \href{http://dx.doi.org/10.1098/rspa.1998.0164}{Proceedings of
  the Royal Society of London. Series A: Mathematical, Physical and Engineering
  Sciences {\bfseries 454} no.~1969, (Jan, 1998) 339–354}.
  \url{http://dx.doi.org/10.1098/rspa.1998.0164}.

\bibitem{nielsen2002quantum}
M.~A. Nielsen and I.~Chuang, ``Quantum computation and quantum information,''
  2002.

\bibitem{Gattringer:2010zz}
C.~Gattringer and C.~B. Lang,
  \href{http://dx.doi.org/10.1007/978-3-642-01850-3}{{\em {Quantum
  chromodynamics on the lattice}}}, vol.~788.
\newblock Springer, Berlin, 2010.

\bibitem{Izubuchi:2008mu}
T.~Izubuchi, S.~Aoki, K.~Hashimoto, Y.~Nakamura, T.~Sekido, and G.~Schierholz,
  ``{Dynamical QCD simulation with theta terms},''
  \href{http://dx.doi.org/10.22323/1.042.0106}{PoS {\bfseries LATTICE2007}
  (2007) 106}, \href{http://arxiv.org/abs/0802.1470}{{\ttfamily arXiv:0802.1470
  [hep-lat]}}.

\bibitem{Aarts:2015tyj}
G.~Aarts, ``{Introductory lectures on lattice QCD at nonzero baryon number},''
  \href{http://dx.doi.org/10.1088/1742-6596/706/2/022004}{J. Phys. Conf. Ser.
  {\bfseries 706} no.~2, (2016) 022004},
  \href{http://arxiv.org/abs/1512.05145}{{\ttfamily arXiv:1512.05145
  [hep-lat]}}.

\bibitem{Alexandru:2016gsd}
A.~Alexandru, G.~Basar, P.~F. Bedaque, S.~Vartak, and N.~C. Warrington,
  ``{Monte Carlo Study of Real Time Dynamics on the Lattice},''
  \href{http://dx.doi.org/10.1103/PhysRevLett.117.081602}{Phys. Rev. Lett.
  {\bfseries 117} no.~8, (2016) 081602},
  \href{http://arxiv.org/abs/1605.08040}{{\ttfamily arXiv:1605.08040
  [hep-lat]}}.

\bibitem{Takeda:2019idb}
S.~Takeda, ``{Tensor network approach to real-time path integral},''
  \href{http://dx.doi.org/10.22323/1.363.0033}{PoS {\bfseries LATTICE2019}
  (2019) 033}, \href{http://arxiv.org/abs/1908.00126}{{\ttfamily
  arXiv:1908.00126 [hep-lat]}}.

\bibitem{feynman1982simulating}
R.~P. Feynman, ``Simulating physics with computers,'' Int. J. Theor. Phys
  {\bfseries 21} no.~6/7, (1982) .

\bibitem{kitaev1995quantum}
A.~Y. Kitaev, ``Quantum measurements and the abelian stabilizer problem,''
  arXiv preprint quant-ph/9511026 (1995) .

\bibitem{shor1994algorithms}
P.~W. Shor, ``Algorithms for quantum computation: discrete logarithms and
  factoring,'' in {\em Proceedings 35th annual symposium on foundations of
  computer science}, pp.~124--134, Ieee.
\newblock 1994.

\bibitem{2000quant.ph..5055B}
G.~{Brassard}, P.~{Hoyer}, M.~{Mosca}, and A.~{Tapp}, ``{Quantum Amplitude
  Amplification and Estimation},'' arXiv e-prints (May, 2000)
  quant--ph/0005055, \href{http://arxiv.org/abs/quant-ph/0005055}{{\ttfamily
  arXiv:quant-ph/0005055 [quant-ph]}}.

\bibitem{Harrow_2009}
A.~W. Harrow, A.~Hassidim, and S.~Lloyd, ``Quantum algorithm for linear systems
  of equations,''
  \href{http://dx.doi.org/10.1103/physrevlett.103.150502}{Physical Review
  Letters {\bfseries 103} no.~15, (Oct, 2009) }.
  \url{http://dx.doi.org/10.1103/PhysRevLett.103.150502}.

\bibitem{Childs_2017}
A.~M. Childs, R.~Kothari, and R.~D. Somma, ``Quantum algorithm for systems of
  linear equations with exponentially improved dependence on precision,''
  \href{http://dx.doi.org/10.1137/16m1087072}{SIAM Journal on Computing
  {\bfseries 46} no.~6, (Jan, 2017) 1920–1950}.
  \url{http://dx.doi.org/10.1137/16M1087072}.

\bibitem{LCU_2012}
A.~M. Childs and N.~Wiebe, ``Hamiltonian simulation using linear combinations
  of unitary operations,'' Quantum Info. Comput. {\bfseries 12} no.~11–12,
  (Nov., 2012) 901–924.

\bibitem{sprep_Cirac}
Y.~Ge, J.~Tura~Brugués, and J.~Cirac, ``Faster ground state preparation and
  high-precision ground energy estimation with fewer qubits,''
  \href{http://dx.doi.org/10.1063/1.5027484}{Journal of Mathematical Physics
  {\bfseries 60} (02, 2019) 022202},
  \href{http://arxiv.org/abs/1712.03193}{{\ttfamily arXiv:1712.03193
  [quant-ph]}}.

\bibitem{Lin_2020}
L.~Lin and Y.~Tong, ``Near-optimal ground state preparation,''
  \href{http://dx.doi.org/10.22331/q-2020-12-14-372}{Quantum {\bfseries 4}
  (Dec, 2020) 372}. \url{http://dx.doi.org/10.22331/q-2020-12-14-372}.

\bibitem{cosine_2006}
W.~{van Dam}, G.~M. {D'Ariano}, A.~{Ekert}, C.~{Macchiavello}, and M.~{Mosca},
  ``{Optimal Quantum Circuits for General Phase Estimation},''
  \href{http://dx.doi.org/10.1103/PhysRevLett.98.090501}{\prl {\bfseries 98}
  no.~9, (Mar., 2007) 090501},
  \href{http://arxiv.org/abs/quant-ph/0609160}{{\ttfamily
  arXiv:quant-ph/0609160 [quant-ph]}}.

\bibitem{Ba_uls_2019}
M.~C. Bañuls, K.~Cichy, Y.-J. Kao, C.-J.~D. Lin, Y.-P. Lin, and D.~T.-L. Tan,
  ``Phase structure of the ( 1+1 )-dimensional massive thirring model from
  matrix product states,''
  \href{http://dx.doi.org/10.1103/physrevd.100.094504}{Physical Review D
  {\bfseries 100} no.~9, (Nov, 2019) }.
  \url{http://dx.doi.org/10.1103/PhysRevD.100.094504}.

\bibitem{Harris78onthe}
F.~J. Harris, ``On the use of windows for harmonic analysis with the discrete
  fourier transform,'' in {\em Proc. IEEE}, pp.~51--83.
\newblock 1978.

\bibitem{blackman1958measurement}
R.~B. Blackman and J.~W. Tukey, ``The measurement of power spectra from the
  point of view of communications engineering—part i,'' Bell System Technical
  Journal {\bfseries 37} no.~1, (1958) 185--282.

\bibitem{whittaker_e_t_1915_1428702}
E.~T. Whittaker, ``{XVIII.—On the Functions which are represented by the
  Expansions of the Interpolation-Theory},'' Jan., 1915.
\newblock \url{https://doi.org/10.1017/s0370164600017806}.

\bibitem{inbook}
V.~Kotel’nikov, {\em On the Transmission Capacity of the “Ether” and Wire
  in Electrocommunications},
  \href{http://dx.doi.org/10.1007/978-1-4612-0143-4_2}{pp.~27--45}.
\newblock 01, 2001.

\bibitem{Shannon_sampling}
C.~E. {Shannon}, ``Communication in the presence of noise,''
  \href{http://dx.doi.org/10.1109/JRPROC.1949.232969}{Proceedings of the IRE
  {\bfseries 37} no.~1, (1949) 10--21}.

\bibitem{ipol.2019.273}
T.~Briand, ``{Trigonometric Polynomial Interpolation of Images},''
  \href{http://dx.doi.org/10.5201/ipol.2019.273}{{Image Processing On Line}
  {\bfseries 9} (2019) 291--316}.

\bibitem{Wilkinson_EVP}
J.~H. Wilkinson, {\em The Algebraic Eigenvalue Problem}.
\newblock Oxford University Press, Inc., USA, 1988.

\bibitem{YiCrosson_prod_pert}
C.~{Yi} and E.~{Crosson}, ``{Spectral Analysis of Product Formulas for Quantum
  Simulation},'' arXiv e-prints (Feb., 2021) arXiv:2102.12655,
  \href{http://arxiv.org/abs/2102.12655}{{\ttfamily arXiv:2102.12655
  [quant-ph]}}.

\bibitem{Berry_2015_v1}
D.~W. Berry, A.~M. Childs, and R.~Kothari, ``Hamiltonian simulation with nearly
  optimal dependence on all parameters,''
  \href{http://dx.doi.org/10.1109/focs.2015.54}{2015 IEEE 56th Annual Symposium
  on Foundations of Computer Science (Oct, 2015) }.
  \url{http://dx.doi.org/10.1109/FOCS.2015.54}.

\bibitem{Berry_2015_v2}
D.~W. Berry, A.~M. Childs, R.~Cleve, R.~Kothari, and R.~D. Somma, ``Simulating
  hamiltonian dynamics with a truncated taylor series,''
  \href{http://dx.doi.org/10.1103/physrevlett.114.090502}{Physical Review
  Letters {\bfseries 114} no.~9, (Mar, 2015) }.
  \url{http://dx.doi.org/10.1103/PhysRevLett.114.090502}.

\bibitem{Low_2017}
G.~H. Low and I.~L. Chuang, ``Optimal hamiltonian simulation by quantum signal
  processing,''
  \href{http://dx.doi.org/10.1103/physrevlett.118.010501}{Physical Review
  Letters {\bfseries 118} no.~1, (Jan, 2017) }.
  \url{http://dx.doi.org/10.1103/PhysRevLett.118.010501}.

\bibitem{Low2019hamiltonian}
G.~H. Low and I.~L. Chuang, ``Hamiltonian {S}imulation by {Q}ubitization,''
  \href{http://dx.doi.org/10.22331/q-2019-07-12-163}{{Quantum} {\bfseries 3}
  (July, 2019) 163}. \url{https://doi.org/10.22331/q-2019-07-12-163}.

\bibitem{Berry_2006}
D.~W. Berry, G.~Ahokas, R.~Cleve, and B.~C. Sanders, ``Efficient quantum
  algorithms for simulating sparse hamiltonians,''
  \href{http://dx.doi.org/10.1007/s00220-006-0150-x}{Communications in
  Mathematical Physics {\bfseries 270} no.~2, (Dec, 2006) 359–371}.
  \url{http://dx.doi.org/10.1007/s00220-006-0150-x}.

\bibitem{Suzuki1991GeneralTO}
M.~Suzuki, ``General theory of fractal path integrals with applications to
  many‐body theories and statistical physics,'' Journal of Mathematical
  Physics {\bfseries 32} (1991) 400--407.

\bibitem{Low_Interp}
G.~{Hao Low}, V.~{Kliuchnikov}, and N.~{Wiebe}, ``{Well-conditioned
  multiproduct Hamiltonian simulation},'' arXiv e-prints (July, 2019)
  arXiv:1907.11679, \href{http://arxiv.org/abs/1907.11679}{{\ttfamily
  arXiv:1907.11679 [quant-ph]}}.

\bibitem{farhi2014quantum}
E.~Farhi, J.~Goldstone, and S.~Gutmann, ``A quantum approximate optimization
  algorithm,'' \href{http://arxiv.org/abs/1411.4028}{{\ttfamily arXiv:1411.4028
  [quant-ph]}}.

\bibitem{Wecker2015}
D.~Wecker, M.~B. Hastings, and M.~Troyer, ``Progress towards practical quantum
  variational algorithms,''
  \href{http://dx.doi.org/10.1103/PhysRevA.92.042303}{Phys. Rev. A {\bfseries
  92} (Oct, 2015) 042303}.
  \url{https://link.aps.org/doi/10.1103/PhysRevA.92.042303}.

\bibitem{farhi2019quantum}
E.~Farhi and A.~W. Harrow, ``Quantum supremacy through the quantum approximate
  optimization algorithm,'' \href{http://arxiv.org/abs/1602.07674}{{\ttfamily
  arXiv:1602.07674 [quant-ph]}}.

\bibitem{Zhou2019}
L.~Zhou, S.-T. Wang, S.~Choi, H.~Pichler, and M.~D. Lukin, ``Quantum
  approximate optimization algorithm: Performance, mechanism, and
  implementation on near-term devices,''
  \href{http://dx.doi.org/10.1103/PhysRevX.10.021067}{Phys. Rev. X {\bfseries
  10} (Jun, 2020) 021067}.

\bibitem{Ho_2019}
W.~W. Ho and T.~H. Hsieh, ``Efficient variational simulation of non-trivial
  quantum states,''
  \href{http://dx.doi.org/10.21468/scipostphys.6.3.029}{SciPost Physics
  {\bfseries 6} no.~3, (Mar, 2019) }.
  \url{http://dx.doi.org/10.21468/SciPostPhys.6.3.029}.

\end{thebibliography}
\providecommand{\href}[2]{#2}\begingroup\raggedright\endgroup

\end{document}